\documentclass[aos,MSNbibl,seceqn,rotating,dvips]{arximspdf}
\usepackage{mathbh,dcolumn}

\doi{10.1214/12-AOS1007} 
\volume{40}
\issue{3}
\pubyear{2012}
\firstpage{1315}
\lastpage{1345}

\makeatletter
\newcolumntype{d}[1]{D{.}{.}{#1}}
\newcommand{\tr}{\operatorname{tr}}
\newcommand{\diag}{\operatorname{diag}}
\newcommand{\sgn}{\operatorname{sgn}}

\newproclaim{remark}{Remark}[section]
\newtheorem{prop}{Proposition}[section]
\newtheorem{teo}{Theorem}[section]
\newproclaim{ex}{Example}[section]
\newtheorem{cor}{Corollary}[section]
\makeatother

\begin{document}
\begin{frontmatter}

\title{Multi-objective optimal designs in comparative clinical trials
with covariates: The reinforced doubly adaptive biased coin design}
\runtitle{The reinforced doubly-adaptive biased coin design}

\begin{aug}
\author[A]{\fnms{Alessandro} \snm{Baldi Antognini}\corref{}\ead[label=e1]{a.baldi@unibo.it}}
\and
\author[A]{\fnms{Maroussa} \snm{Zagoraiou}\ead[label=e2]{maroussa.zagoraiou@unibo.it}}
\runauthor{A. Baldi Antognini and M. Zagoraiou}
\affiliation{University of Bologna}
\address[A]{Department of Statistical Sciences\\
University of Bologna\\
Via Belle Arti 41, 40126\\
Bologna\\
Italy\\
\printead{e1}\\
\phantom{E-mail: }\printead*{e2}} 
\end{aug}

\received{\smonth{3} \syear{2012}}

%
\begin{abstract}
The present paper deals with the problem of allocating patients to two
competing treatments in the presence of covariates or prognostic
factors in order to achieve a good
trade-off among ethical concerns, inferential precision and randomness
in the treatment allocations.
In particular we suggest a multipurpose design methodology that
combines efficiency and ethical gain when the linear homoscedastic
model with both treatment/covariate
interactions and interactions among covariates is adopted. The ensuing
compound optimal allocations of the treatments depend on the covariates
and their distribution on the population of interest, as well as on the
unknown parameters of the model. Therefore, we introduce the reinforced
doubly adaptive biased coin design, namely a general class of
covariate-adjusted response-adaptive procedures that includes both
continuous and discontinuous randomization functions, aimed to target
any desired allocation proportion.
The properties of this proposal are described both theoretically and
through simulations.
\end{abstract}

%

\end{frontmatter}
%

\section{Introduction}
In the medical profession physicians are expected to act in the best
interests of each patient
under their care, but in clinical trials the patients are statistical
units in an experiment, and the demands of individual care and
experimental information often come into conflict. Thus the ensuing
ethical problem is how to
balance the welfare of the patients in the trial against a possible
knowledge gain that will
improve the care of future patients. In experimental medicine it is
commonly believed
that randomized trials are the answer, especially in the case of Phase
III clinical trials, where the aim is to compare the efficacy of two
available treatments and patients are sequentially randomized to one of them.
In this context, several design methodologies have been recently
proposed in order to derive suitable target allocations of the
treatments that represent a compromise between ethical demands and
inferential goals \cite{Bag2010,Ros01,Tym07}. These targets depend in
general on the unknown model parameters and can be implemented by
adopting suitable response-adaptive procedures, such as the sequential
maximum likelihood design \cite{Mel01}, the doubly adaptive biased coin
design \cite{Hu04} and the efficient randomized-adaptive design (ERADE)
\cite{Hu09}, converging to them.

An additional peculiarity of Phase III trials is that they usually
involve some set of important prognostic factors or covariates.
The role of these factors may be crucial in order to derive correct
inferential conclusions about the treatment effects
and this is one of the reasons for which taking into account the
covariates has now become of primary importance not only from an
inferential point of view but also from a design perspective (for
a~recent literature \cite{ATK02,Baz11,Ros08,Sha10}).

In the last decade there has been a growing statistical interest in the
topic of adaptive designs adjusted for covariates and, in particular,
in covariate-adjusted response-adaptive (CARA) randomization methods;
see \cite{Hu07,RVA01,Zha07}.
This is a class of sequential allocation procedures that modifies the
probabilities of treatment assignments on the basis of the available
information---that is, earlier responses and allocations, past
covariate profiles and the covariate information of the present
patient---with the aim of skewing the allocations towards the treatment
that appears to be superior or, more generally, of converging to a
desired target that should incorporate inferential demands related to
optimal inference about the treatment effects and, eventually, ethical concerns.

Even if in the presence of prognostic factors the inferential methods,
as well as the ethical goals, change on the basis of the nature of the
responses and the covariates, the design literature has essentially
focused on the simplified scenario of absence of treatment/covariate
interactions; see, for instance, \cite{Atk05a,Ban01,Ban07}. In this
context, the relative performance of the treatments is the same for
every subject's profile, so that the ethical demand simply consists of
allocating the best treatment to as many patients as possible,
independently on their
covariates. This also explains the absence of methodological proposals
through which one can derive target allocations that, by incorporating
both inferential and ethical considerations, depend on the covariates.
Whereas, in the more complex scenario of treatment/covariate
interactions, the covariates play a fundamental role also from an
ethical viewpoint, since the superiority/inferiority of a given
treatment, as well as the discrepancy between the treatment effects,
depend on the subject's profiles.

Therefore, one of the primary aims of the present paper consists of
analyzing, from a design perspective, the linear model with both
treatment/covari\-ate
interactions and interactions among covariates, in terms of ethical
impact as well as inferential efficiency. After deriving the analytical
expressions of the most popular information criteria, we propose a
multipurpose design strategy based on a compound optimization approach,
that combines
inferential precision and ethical gain by means of flexible weights,
which can be fixed a~priori by the experimenter, or they may be
functions of the unknown model parameters.

This multipurpose criterion leads to a locally optimal allocation which
depends, in general, on the covariates and their population
distribution, as well as on the unknown model parameters, and allows to
promote for every profile a suitable compromise between information and
ethical demands. Furthermore,
we introduce the reinforced doubly adaptive biased coin design (RDBCD),
namely a new class of CARA procedures that generalizes some earlier
works \cite{Atk82,Zha07} and also extends several procedures proposed
in the literature, such as the doubly adaptive biased coin design and
the ERADE, to the covariate setting.
The RDBCD, which admits both continuous and discontinuous randomization
functions, can target any desired allocation proportion, allowing also
to force closeness to the chosen target in an appropriate way, while
maintaining randomization. We show, both theoretically and through
simulations, that the proposed procedure has desirable properties,
asymptotically and, in particular, for small samples.

The paper is structured as follows. Starting from the notation
in Section~\ref{s2}, Sections~\ref{s3} and~\ref{s4} deal with the optimal designs for
inference and ethics, respectively. Section~\ref{s5} describes the combined
approach, while the properties of the ensuing optimal compound target
are discussed in Section~\ref{s6}. In Section~\ref{s7} we introduce the RDBCD,
showing its asymptotic properties as well as the asymptotic inference
related to the adoption of such a procedure. Section~\ref{s8} deals with some
finite sample comparisons between our proposal and some of the
fundamental procedures proposed in the literature. We end the paper
with a brief discussion in Section~\ref{s9}.

Motivated by the clinical practice, the present paper takes into
account categorical covariates, since in the large majority of real
Phase III clinical trials the prognostic factors are polytomous and,
even if quantitative, they are often categorized by adopting suitable
thresholds. Furthermore, for ease of notation we deal with just two
covariates, but the extension to the case of several factors is
straightforward (see \cite{Baz12}).

\section{The linear model with covariates}\label{s2}
Let $A$ and $B$ be two competing treatments. We suppose that for
each subject entering the trial we observe a vector
$\mathbf{Z}$ of concomitant categorical variables.
Moreover, we assume the covariates to be
random, that is, they are not under the experimenters' control when the
subjects turn up for the trial, but they can
be measured before assigning a treatment. Then the treatments are
assigned according to a given randomization rule, with $\delta$ an
indicator variable such that $\delta=1$ or $0$ if the subject is
assigned to $A$ or
$B$, respectively, and an outcome~$Y$ is observed. Conditionally on the
covariates and the treatments, patients' responses are assumed to be
independent.
A common model for the response that accounts for treatment/covariate
interactions is the following linear homoscedastic model:
%
\begin{eqnarray}\label{linmod4}
E(Y_{i}) &=&\delta_{i}\mu_{A}+(1-\delta_{i})\mu_{B}+\mathbf{f}(%
\mathbf{z}_{i})^{t}\bigl( \delta_{i}\boldsymbol{\beta
}_{A}+(1-\delta
_{i})\boldsymbol{\beta}_{B}\bigr),\nonumber
\\[-8pt]
\\[-8pt]
V(Y_{i}) &=&\sigma^{2}, \qquad i=1,\ldots,n, \nonumber
\end{eqnarray}
where $Y_{i}$ is the outcome of the $i$th subject, $\mu_{A}$ and $\mu
_{B}$ are the baseline treatment effects, $\mathbf{f}(\cdot)$ is a
known vector function, $\mathbf{z}_{i}$ is the
vector of covariates observed on the $i$th individual and
$\boldsymbol{\beta}_{A},\boldsymbol{\beta}_{B}$ are $p$-dim vectors of
possibly different regression parameters related to $A$ and $B$, respectively.
Under this model $\mu_{A}, \mu_{B}, \boldsymbol{\beta}_{A}$ and
$\boldsymbol{\beta}_{B}$ are of interest since
the relative performance of the treatments depends on the
patient's covariates. Indeed, for any given covariate profile $\mathbf
{z}_i$, we obtain
%
\begin{eqnarray}\label{etica2}
\theta(\mathbf{z}_i)&=& E(Y_i \mid\delta_i =1,\mathbf{Z}_i=\mathbf
{z}_i)-E(Y_i \mid\delta_i=0,\mathbf{Z}_i=\mathbf{z}_i)\nonumber
\\[-8pt]
\\[-8pt]
&=& \mu_{A}-\mu_{B}+\mathbf{f}(\mathbf{z}_i)^{t}( \boldsymbol{\beta
}_{A}-\boldsymbol{\beta}_{B})=\alpha+\mathbf{f}(\mathbf
{z}_i)^{t}\boldsymbol{\tau}.\nonumber
\end{eqnarray}
After $n$ assignments, let $\mathbf{F}=( \mathbf{f}(\mathbf{z}%
)^{t})_{n\times p} $, $\boldsymbol{\delta}_n=(\delta
_{1},\ldots,\delta_{n})^{t}$, $\boldsymbol{\Delta}_n=\diag(\boldsymbol{\delta}_n)$ and
$\mathbf{X}=[\boldsymbol{\delta}_n\dvtx\mathbf{1}_n-\boldsymbol{\delta
}_n\dvtx\boldsymbol{\Delta}_n \mathbf{F}\dvtx(\mathbf{I}_n-\boldsymbol
{\Delta}_n
)\mathbf{F} ]$, where $\mathbf{1}_n$ and $\mathbf{I}_n$ are the $n$-dim
vector of ones and the $n$-dim identity matrix, respectively. Moreover,\vspace*{2pt}
let $\boldsymbol{\hat{\gamma}}_n=(\hat{\mu} _{A},\hat{\mu}
_{B},\boldsymbol{\hat{\beta} }{}_{A}^{t},\vspace*{2pt}
\boldsymbol{\hat{\beta}}{}_{B}^t )^{t}$ be the
least
square estimator of $\boldsymbol{\gamma}=(\mu_{A},\mu
_{B},\boldsymbol
{\beta}_{A}^t,
\boldsymbol{\beta}_{B}^t)^{t}$, then if $(\mathbf{X}^{t}\mathbf
{X})^{-1}$ exists, the variance--covariance matrix is
\[
V(\boldsymbol{\hat{\gamma}}_n)=\sigma^{2}(\mathbf{X}^{t}\mathbf{X})^{-1}=
{n}^{-1}\sigma^{2}\mathbf{M}^{-1},
\]
where $\mathbf{M}$ is the $(2+2p)$-dim average (\emph{per observation})
information matrix
%
\begin{equation}\label{MconInter}
\mathbf{M}=\frac{1}{n}\left(
\begin{array} {c@{\quad}c@{\quad}c@{\quad}c}
\displaystyle\sum_{i=1}^n\delta_i & 0 & \boldsymbol{\delta
}_n^{t}\mathbf{F} &\mathbf{0}_{1\times p} \\
0 & n-\displaystyle\sum_{i=1}^n\delta_i & \mathbf{0}_{1\times p} & (
\mathbf{1}_n-\boldsymbol{\delta}_n)^{t}\mathbf{F} \\
\mathbf{F}^{t}\boldsymbol{\delta}_n & \mathbf{0}_{p\times1} &
\mathbf{F}^{t}\boldsymbol{\Delta}_n
\mathbf{F} & \mathbf{0}_{p\times p} \\
\mathbf{0}_{p\times1} & \mathbf{F}^{t}(\mathbf{1}_n-\boldsymbol
{\delta}_n) &\mathbf{0}_{p\times p}
& \mathbf{F}^{t}(\mathbf{I}_n-\boldsymbol{\Delta}_n )\mathbf{F}
\end{array}
\right).
\end{equation}
\begin{remark}
In the absence of treatment/covariate interactions $\boldsymbol{\beta
}_{A}=\boldsymbol{\beta}_{B}=\boldsymbol{\beta}$, that is,
$\boldsymbol{\tau}=\mathbf{0}$, and the homoscedastic model (\ref
{linmod4}) becomes
%
\begin{equation}\label{linmod}
E(Y_{i}) =\delta_{i}\mu_{A}+(1-\delta_{i})\mu_{B}+\mathbf
{f}(\mathbf
{z}_{i})^{t}\boldsymbol{\beta}, \qquad i=1,\ldots,n.
\end{equation}
In this case, it is customary to regard $\boldsymbol{\beta}$ as a
nuisance parameter, since from~(\ref{etica2}) $\theta(\mathbf
{z}_i)=\mu
_{A}-\mu_{B}$ for any given covariate profile $\mathbf{z}_i$, so that
the inferential interest typically lies in estimating $\mu_{A}$ and
$\mu_{B}$, or $\mu_{A}-\mu_{B}$, as precisely as possible.
\end{remark}

\section{Inferential optimality and balanced designs}\label{s3}
In order to avoid cumbersome notation, from now on we assume, without
loss of generality, only two categorical
covariates, that is, $\mathbf{Z}=(T,W)$. Suppose that $T$ is
categorized into levels $t_0,t_1,\ldots,t_J$ and let $w_0,w_1,\ldots
,w_L$ be the levels
of $W$, so
that $T$ and $W$ can be represented by a $J$-dimensional vector
$\mathbb{T}$ and a $L$-dimensional vector $\mathbb{W}$ of
dummy variables, respectively, where $t_0$ and~$w_0$ are the reference
categories. Assume that
$\{Z_i, i\geq1\}$ is a sequence of i.i.d.
random vectors, where each $Z_i$ is distributed in the population according
to $\Pr\{Z_i=(t_j,w_l)\}=p(j,l)> 0$ ($j=0,\ldots,J;l=0\ldots,L$),
where \mbox{$\sum_{j=0}^{J}\sum_{l=0}^{L}p(j,l)=1$}.
Moreover, in order to account for the general situation of both
treatment/\break covariate interactions and interaction among covariates in
the rest of the paper, we let
%
\begin{equation}\label{f}
\mathbf{f}^{t}(\mathbf{z})=(\mathbb{T}^{t} ,
\mathbb{W}^{t} ,\mathbb{T}^{t}\otimes
\mathbb{W}^{t}),
\end{equation}
that is, $\mathbf{f}(\cdot)$ is the $p$-dim vector including all
interaction effects, with $p=J+L+J\cdot L$.

At the end of a trial with $n$ assignments, let $N_n(j,l)=\sum
_{i=1}^{n}\mathbh{1}_{\{Z_i=(t_j,w_l)\}}$ be the number of subjects
within the
stratum $(t_j,w_l)$, where $\mathbh{1}_{\{\cdot\}}$ represents the
indicator function, $\widetilde{N}_n(j,l)=\sum_{i=1}^{n}\delta
_i\mathbh{1}_{\{Z_i=(t_j,w_l)\}}$ the number of allocations to $A$ within this
stratum and $\pi_n(j,l)$
the corresponding proportion, that is, $\pi
_n(j,l)=N_n(j,l)^{-1}\widetilde{N}_n(j,l)$, for any $j=0,\ldots,J$ and
$l=0,\ldots,L$. Moreover, let $\mathbf{N}_n=\{N_n(j,l)\dvtx j=0,\ldots
,J;l=0,\ldots,L\}$ and $\boldsymbol{\pi}_n=\{\pi_n(j,l)\dvtx j=0,\ldots
,J;l=0,\ldots,L\}$.

Adopting model (\ref{linmod4}), the design for
optimal inference consists in allocating the treatments so
as to minimize one of the following criteria:
\begin{enumerate}[(C4)]
\item[C1] $ \det V(\boldsymbol{\hat{\gamma}}_n)= \det(n^{-1}\sigma
^{2}\mathbf{M}^{-1})$;
\item[C2] $ \det V {\boldsymbol{\hat{\beta}}_{A}\choose
\boldsymbol{\hat{\beta}}_{B}}
 = \det( n^{-1}\sigma^{2}\mathbf{D}^{t}\mathbf{M}^{-1}\mathbf{D}),
\mbox{ where } \mathbf{D}^{t}=(\mathbf{0}_{2p\times2}\dvtx \mathbf
{I}_{2p})$;
\item[C3] $ \tr V(\boldsymbol{\hat{\gamma}}_n)= \tr(n^{-1}\sigma
^{2}\mathbf{M}^{-1})$;
\item[C4] $ \tr V{\boldsymbol{\hat{\beta}}_{A} \choose
\boldsymbol{\hat{\beta}}_{B}}
 =\tr( n^{-1}\sigma^{2}\mathbf{D}^{t}\mathbf{M}^{-1}\mathbf{D})$;
\item[C5] $ \tr V( \boldsymbol{\hat{\beta}}_{A}-\boldsymbol{\hat
{\beta}}_{B}) =\tr(n^{-1}\sigma^{2}\mathbf{E}^{t}\mathbf{M}^{-1}\mathbf{E}),
\mathbf{E}^{t}=(\mathbf{0}_{p\times2} \dvtx \mathbf{I}_{p} \dvtx -\mathbf{I}_{p})$
\end{enumerate}
with $\mathbf{M}^{-1}$ replaced by the Moore--Penrose inverse, if
needed. It is easy to check that C1--C5 are convex
functions of $\mathbf{M}$, invariant with respect to permutations of
the bottom two block rows and the
two right-hand block columns of $\mathbf{M}$.

For given covariates, the jointly balanced design
%
\begin{equation}\label{InfOpti}
\boldsymbol{\pi}_I^{\ast}=\{\pi_I^{\ast}(j,l)=1/2, \mbox{ for any
}j=0,\ldots,J \mbox{ and } l=0,\ldots,L\},
\end{equation}
is optimal for model (\ref{linmod4}) with respect to any criterion
$\Phi_I$ of the information matrix $\mathbf{M}$, which is convex and
invariant w.r.t. permutations of the bottom two block rows and the
two right-hand block columns, as well as the first two rows and
columns. To see this, it is straightforward to check that
assuming~(\ref{InfOpti}) the ensuing information matrix
%
\begin{equation}\label{M_InterOpt}
\mathbf{M}^{\ast}=\frac{1}{2n} \left(
\begin{array} {c@{\quad}c@{\quad}c@{\quad}c}
n & 0 & \mathbf{1}_n^{t}\mathbf{F} & \mathbf{0}_{1\times p}\\
0 & n & \mathbf{0}_{1\times p} & \mathbf{1}_n^{t}\mathbf{F} \\
\mathbf{F}^{t}\mathbf{1}_n & \mathbf{0}_{p\times1} & \mathbf
{F}^{t}\mathbf{F} &\mathbf{0}_{p\times p}\\
\mathbf{0}_{p\times1} & \mathbf{F}^{t}\mathbf{1}_n & \mathbf
{0}_{p\times p} & \mathbf{F}^{t}\mathbf{F}
\end{array}
\right)
\end{equation}
is invariant w.r.t. permutations of the bottom two block rows and the
two right-hand block columns, as well as the first two rows and
columns.
For any information matrix $\mathbf{M}$ of the type
(\ref{MconInter}), by the simultaneous permutation of the first two
rows and two columns as well as the bottom two block rows and the
two right-hand block columns, we get the information matrix $\mathbf
{\tilde{M}}$
corresponding to the design which
switches treatments $A$ and $B$. Clearly $\Phi_I(\mathbf{\tilde
{M}})=\Phi_I(\mathbf{M)}$, $( \mathbf{M}+\mathbf{\tilde{M}})/2=
\mathbf
{M}^{\ast}$
and then by convexity\looseness=1
\[
\Phi_I(\mathbf{M}^{\ast})=\Phi_I\bigl(\tfrac{1}{2}(\mathbf{M}+\mathbf
{\tilde{M}})\bigr) \leq\tfrac{1}{2}[\Phi_I(\mathbf{M)}+\Phi_I(\mathbf{\tilde
{M}})]=\Phi_I (\mathbf{M}).
\]\looseness=0
Note that, independently on the presence or absence of
treatment/covariate interactions, the jointly balanced allocation (\ref
{InfOpti}) is still optimal, even in the absence of interactions among
covariates; see \cite{Baz11} for a detailed discussion.\looseness=1

\begin{prop}\label{mainprop}
Assuming model (\ref{linmod4}) with (\ref{f}), inferential criteria~\emph{C1--C5}
can be simplified as follows:
%
\begin{equation}\label{Dopt}
\hspace*{2pt} \mbox{\emph{C1}:} \quad \det\biggl(
\frac{\sigma^{2}}{n}\mathbf{M}^{-1}\biggr) = \frac{ \sigma^{4+4p} } {
\prod_{j=0}^{J} \prod_{l=0}^{L} \pi_n(j,l) [1-\pi_n(j,l)] N_n(j,l)^2 },
\end{equation}\vspace*{-8pt}
%
\begin{equation}\label{C3_crit}
\qquad\mbox{\emph{C2}:}\quad\det\biggl( \frac{\sigma^{2}}{n}\mathbf{D}^{t}\mathbf
{M}^{-1}\mathbf{D}\biggr) = \frac{( \sum_{i=1}^{n} \delta_i ) ( n-\sum
_{i=1}^{n} \delta_i ) \sigma^{4p} } { \prod_{j=0}^{J}\prod
_{l=0}^{L}\pi
_n(j,l) [1-\pi_n(j,l)] N_n(j,l)^2 },
\end{equation}\vspace*{-8pt}
%
\begin{eqnarray}\label{trgamma}
\hspace*{10pt}\mbox{\emph{C3}:}  \hspace*{8pt} &&\tr\biggl(\frac{\sigma^{2}}{n}\mathbf{M}^{-1}\biggr)
=\sigma^2 \times\Biggl[ \sum_{j=1}^{J} \sum_{l=1}^{L} \frac{1} {N_n(j,l) \pi_n(j,l)[1-\pi
_n(j,l)]}\nonumber\\
&&\qquad \hspace*{76pt} {}+ \sum_{j=1}^{J} \frac{L+1} {N_n(j, 0) \pi_n(j, 0)[1-\pi_n(j,0)]}\nonumber
\\[-8pt]
\\[-8pt]
&&\qquad \hspace*{76pt} {}+ \sum_{l=1}^{L} \frac{J+1} {N_n (0, l) \pi_n(0, l)[1-\pi_n(0,l)]}\nonumber\\
&& \qquad \hspace*{92.5pt} {} + \frac{(J+1) \times(L+1)} {N_n (0, 0)\pi_n(0, 0)[1-\pi_n(0,
0)]}\Biggr].\nonumber
\end{eqnarray}
Furthermore, criterion \emph{C4} coincides with \emph{C5} and
is given by
%
\begin{eqnarray}\label{traceC4}\qquad
\tr\biggl( \frac{\sigma^{2}}{n}\mathbf{D}^{t}\mathbf{M}^{-1}\mathbf{D}\biggr)
&= &
\tr\biggl( \frac{\sigma^{2}}{n}\mathbf{E}^{t}\mathbf{M}^{-1}\mathbf
{E}\biggr)\nonumber
\\[-8pt]
\\[-8pt]
&=&\tr\biggl(\frac{\sigma^{2}}{n}\mathbf{M}^{-1}\biggr)-\frac{\sigma^2}{N_n (0,
0)\pi_n(0, 0)[1-\pi_n(0, 0)]}.\nonumber
\end{eqnarray}
\end{prop}
\begin{pf}
See Appendix \ref{a1}.
\end{pf}
\begin{remark}
Contrary to C1 and C2, from (\ref{trgamma}) and
(\ref{traceC4}) it is easy to see that the trace criteria C3--C5 treat the covariate profiles in a different way due
to the nature of the OLS estimators and the fact that these criteria
correspond to the minimization of the mean variance of the estimators
of the parameters of interest without taking into account their
covariance structure.
\end{remark}

Note that C1--C5 depend on the design only through
the allocation vector~$\boldsymbol{\pi}$; it is also straightforward to
check that the above criteria are strictly convex in $\boldsymbol{\pi}$
and will be minimized by (\ref{InfOpti}) independently on the
covariates. However, the loss of inferential precision expressed by
C1--C5 is random since it depends on the number of
subjects within the different strata $\mathbf{N}_n$, that is, $\Phi
_I=\Phi_I(\boldsymbol{\pi}_n, \mathbf{N}_n)$.
Therefore, in order to remove the effect due to the random covariates,
from now on we take into
account the loss of inferential precision induced by the design
%
\begin{equation}\label{expcritinfo}
\widetilde{\Phi}_I (\boldsymbol{\pi}_n)=E_{\mathbf{Z}}[\Phi_I
(\boldsymbol{\pi}_n,\mathbf{N}_n)].
\end{equation}

\section{Optimal design for ethics}\label{s4}
From an ethical viewpoint, a natural demand consists in an overall
benefit for the entire sample of patients involved in the trial,
for instance maximizing for any given sample size the percentage of patients
who receive the best treatment.
This make sense if and only if the treatment effects are different; otherwise
there is no longer a worse treatment, stressing that the
comparative experiment degenerates to just one treatment.

Assuming model (\ref{linmod4}), at each stratum $(t_j,w_l)$ the
superiority/inferiority of $A$ or $B$ depends only on the sign of
$\theta$ in (\ref{etica2}), and from now on we let for simplicity
$\theta(j,l)$. Assuming ``the-larger-the-better'' scenario, for each
subject with covariate
profile $(t_j,w_l)$ the allocation will be made to the superior treatment
if $\delta_i\mathbh{1}_{\{\theta(j,l)>0\}}+(1-\delta_i)\mathbh{1}_{\{
\theta(j,l)<0\}}$;
otherwise, if \mbox{$\theta(j,l)=0$} the two treatment arms collapse and
all the allocations are equivalent (i.e., any ethical measurement is no
longer useful). Thus, the percentage of patients assigned to the best
treatment is
\[
\frac{1}{n}\sum_{j=0}^{J}\sum_{l=0}^{L}
N_n(j,l) \biggl\{\frac{1}{2}-\biggl[\frac{1}{2}-\pi_n(j,l)\biggr]\sgn(\theta(j,l)) \biggr\},
\]
where $\sgn(x)$ represents the sign of $x$.
However, from (\ref{etica2}) the relative performance of the treatments
depends on the
subject's covariates, so that a~reasonable ethical measure is
%
\begin{eqnarray}\label{cretica2}
&&\Phi_E(\boldsymbol{\pi}_n,\mathbf{N}_n)\nonumber
\\[-8pt]
\\[-8pt]
&& \qquad=\frac{1}{n}\sum_{j=0}^{J}\sum_{l=0}^{L}
N_n(j,l)|\theta(j,l)|\biggl \{\frac{1}{2}-\biggl[\frac{1}{2}-\pi_n(j,l)\biggr]\sgn
(\theta
(j,l))\biggr \},\nonumber
\end{eqnarray}
under which every choice is weighed by the relative ethical gain
$|\theta(j,l)| $. Obviously,
criterion (\ref{cretica2}) depends on both the covariate profiles and
the unknown parameters of the model and the optimal ethical target,
namely the allocation that assigns all the patients to the better
treatment, is
%
\begin{equation}\label{optet}\qquad
\boldsymbol{\pi}_E^*=\bigl\{\pi_E^*(j,l)=\mathbh{1}_{ \{ \theta(j,l) >0
\} }
\mbox{ for any }j=0,\ldots,J \mbox{ and } l=0,\ldots,L\bigr\}.
\end{equation}

\begin{remark}
In the absence of treatment/covariate interactions, $\theta
(j,l)=\alpha
$ for any $j=0,\ldots,J$ and $l=0,\ldots,L$, so that
criterion (\ref{cretica2}) simply becomes
\[
\Phi_E(\boldsymbol{\pi}_n)=\frac{|\alpha|}{2}+ \Biggl( \frac{1}{n}\sum
_{i=1}^{n} \delta_i -\frac{1}{2}\Biggr) \alpha,
\]
that depends on the design only through the total proportion of
assignments to $A$.
Thus, under (\ref{linmod}) the percentage of allocations to the
best treatment does not depend on the covariates, which are irrelevant
from an ethical viewpoint,
so that the optimal ethical target is $\pi_E^*(j,l)=\mathbh{1}_{\{
\alpha
>0 \}}$ for any~$(j,l)$.
\end{remark}

Analogously to (\ref{expcritinfo}),
in order to remove the random effect due to covariates from now on we
adopt as ethical criterion $\widetilde{\Phi}_E(\boldsymbol{\pi
}_n)=E_{\mathbf{Z}}[\Phi_E (\boldsymbol{\pi}_n,\mathbf{N}_n)]$
given by
%
\begin{equation}\label{expcriteth}
\quad\widetilde{\Phi}_E(\boldsymbol{\pi}_n)=\sum_{j=0}^{J}\sum_{l=0}^{L}
p(j,l)|\theta(j,l)|\biggl \{\frac{1}{2}-\biggl[\frac{1}{2}-\pi_n(j,l)\biggr]\sgn
(\theta(j,l)) \biggr\}.
\end{equation}

\section{The compromise criterion}\label{s5}
In order to obtain a suitable compromise between inferential precision
and ethical demands, there are several possible approaches. Among them,
a trade-off the criteria via a~combined or a~constrained optimization
has, to the best of our knowledge, the strongest theoretical
justification; see, for example, \cite{Bag2010,CLY96,COOK94,Det97,Ros01,Tym07}. For the sake of generality,
we now suggest a~methodology based on the optimization of a~compound
criterion that mediates between information and ethics, since the constrained
optimization approach can be regarded as a special case of this
proposal (as it will be shown in Remark \ref{constrained}).

Note that, for any chosen inferential criterion C1--C5,
$\widetilde{\Phi}_I$ and $\widetilde{\Phi}_E$ are not homogeneous
measures,\vadjust{\goodbreak}
and in order to put them in a comparable scale,
we consider their standardized version, that is,
%
\begin{equation}\label{PsiE}
\Psi_{E}(\boldsymbol{\pi}_n)= \frac{ \widetilde{\Phi
}_{E}(\boldsymbol
{\pi}_n)} {\widetilde{\Phi}_{E}(\boldsymbol{\pi}_E^{\ast})}\quad \mbox{and}\quad
\Psi_{I}(\boldsymbol{\pi}_n)=\frac{\widetilde{\Phi
}_{I}(\boldsymbol{\pi
}_I^{\ast})}{ \widetilde{\Phi}_{I}(\boldsymbol{\pi}_n)},
\end{equation}
where $\boldsymbol{\pi}_I^{\ast}$ in (\ref{InfOpti}) is the optimal
inferential target minimizing $\widetilde{\Phi}^{I}$ and $\boldsymbol
{\pi}_E^{\ast}$ in~(\ref{optet}) maximizes (\ref{expcriteth}) by
assigning all subjects to the best treatment, with $\widetilde{\Phi
}_{E}(\boldsymbol{\pi}_E^{\ast})=E_\mathbf{Z}[|\theta(\mathbf
{z})|]=\sum
_{j=0}^{J} \sum_{l=0}^{L}|\theta(j,l)| p(j,l)$.

Clearly, $\Psi_{E},\Psi_{I}\dvtx [0,1]^{(J+1)\cdot(L+1)}\rightarrow[0,1]$
represent standardized measures of ethical and inferential efficiency,
respectively, that will be maximized.
Therefore, by introducing an ethical weight $\omega$, we let, as a
compromise criterion,
%
\begin{equation}\label{compound criterion}
\Psi_{\omega}(\boldsymbol{\pi}_n) =\omega\biggl\{\frac{1 } {\Psi_{E}
(\boldsymbol{\pi}_n)}\biggr\} +(1-\omega) \biggl\{
\frac{1} {\Psi_{I}(\boldsymbol{\pi}_n)}\biggr\},
\end{equation}
which can be seen as the reciprocal of the weighted harmonic
mean of $\Psi_{E}$ and $\Psi_{I}$; see also \cite{Bag2010,Det97}.

The ethical weight $\omega$ in the compound criterion can be chosen by
the experimenter, with $0\leq\omega<1$ (we assume $\omega\neq1$ in
order to avoid that the ethical impact completely
overcomes the inferential goal). It may be fixed a~priori or could be
modeled as a function of the unknown parameters on the basis of the
given real situation. In the latter case $\omega$ is allowed to depend
on the true state of nature, since it is reasonable to suppose that the more
the effects of the treatments differ, the more important for the patients
are the chances of receiving the best treatment, whereas in the case of
a small difference, which is more
difficult to detect correctly, more emphasis is given on inferential precision.
In particular, under (\ref{linmod4}) the ethical impact depends on the
covariates, and thus, in order to express an overall measure of ethical
risk for the population of interest, from now on we let
$\omega(E_\mathbf{Z}[|\theta(\mathbf{z})|])\dvtx\mathbb{R}^+ \cup\{
0\}
\rightarrow\lbrack0;1)$ to be a
a continuous and increasing function with $\omega(0) \rightarrow0$.

In general, the choice of the weight function depends on the given
applied context. For instance, in Phase III trials the experimenters
have often some information gathered from previous stage trials, and
more attention is usually needed for inference, provided that the
ethical costs are not prohibitive (such as deaths of patients).
Thus, $\omega$ can be chosen to be an S-shaped function as
%
\begin{equation}\label{omegas}
\omega_s(x)=(1+x^{-2})^{-2(s+1)}[2-(1+x^{-2})^{-2}]\qquad \mbox{with } s
\geq0.
\end{equation}
Additionally, since in several clinical situations it is reasonable to
assume that the ethical concern is negligible in the case of small
difference between the treatment performances,
for example, up to a value $\varsigma$ of the overall risk, and then
increases rapidly, we may assume $\omega(x)=0$ for $x\leq\varsigma$,
with $\omega(x)\rightarrow0$ for $x \rightarrow\varsigma^+$, and
$\omega(x)\rightarrow1$ for $x \rightarrow\infty$. Whereas, an
alternative choice for the ethical weight is the cdf of a chi-square
r.v. $\chi^2_{(r)}$, where $\omega$ decreases as the degrees of freedom
$r$ increases.
By fixing small degrees of freedom, the latter choice allows us to
model the ethical impact in order to grow rapidly, even when the
overall ethical risk is moderate.

\begin{teo}
For every inferential criterion \emph{C1--C5}, the compound
criterion $\Psi_{\omega}(\boldsymbol{\pi}_n)$ is a strictly convex
function of $\boldsymbol{\pi}_n$, so there exists a unique target
allocation minimizing (\ref{compound criterion}) which
is the solution of the system of equations
%
\begin{eqnarray}\label{compound target}
&&[\widetilde{\Phi}_{E}(\boldsymbol{\pi}_n)]^2\frac{\partial
\widetilde{\Phi}_{I}(\boldsymbol{\pi}_n)}{\partial\pi_n(j,l)}\nonumber
\\[-8pt]
\\[-8pt]
&&\qquad =\biggl(\frac
{\omega}{1-\omega} \biggr)\widetilde{\Phi}_{E}(\boldsymbol{\pi}_E^{\ast
} ) \widetilde{\Phi}_{I}(\boldsymbol{\pi}_I^{\ast
}) \theta(j,l) p(j,l)\qquad \forall(j,l).\nonumber
\end{eqnarray}
\end{teo}
\begin{pf}
The suggested compound criterion is a linear combination of the
reciprocals of $\Psi_{I}$ and $\Psi_{E}$. Clearly, $[\Psi_{I}]^{-1}$ is
strictly convex; moreover~$\widetilde{\Phi}_{E}$ in (\ref{expcriteth})
is linear, and thus concave, and it is also non-negative, so
that~$\Psi_{E}$ is strictly convex in $\boldsymbol{\pi}_n$. Therefore, criterion
(\ref{compound criterion}) leads to a unique target allocation
satisfying $\nabla\Psi_{\omega}(\boldsymbol{\pi}_n) =\mathbf{0}$, namely
%
\begin{equation}\qquad
\omega\widetilde{\Phi}_{E}(\boldsymbol{\pi}_E^{\ast
})\frac{\partial\{ [ \widetilde{\Phi}_{E}(\boldsymbol{\pi}_n) ]^{-1}
\} }{ \partial\pi_n(j,l)}+\frac{1-\omega}{\widetilde{\Phi
}_{I}(\boldsymbol{\pi}_I^{\ast
})}\biggl\{\frac{\partial\widetilde{\Phi}_{I}(\boldsymbol{\pi
}_n)}{\partial
\pi_n(j,l)}\biggr\}=0\qquad \forall(j,l),
\end{equation}
that leads to (\ref{compound target}) after simple algebra.
\end{pf}

\section{The optimal compound target}\label{s6}
In this section we describe the properties of the allocation
$\boldsymbol{\pi}_{\omega}^*=\{\pi_{\omega}^*(j,l)\dvtx j=0,\ldots
,J;l=0,\ldots,L\}$ that minimizes the compromise criterion $\Psi
_{\omega
}$ in (\ref{compound criterion}), and we shall refer to it as ``optimal
compound target.'' In general, $\boldsymbol{\pi}_{\omega}^*$ depends on
the experimental choice of the inferential criterion $\widetilde{\Phi
}_{I}$ and the ethical weight $\omega$, as well as on the true state of
the nature, that is, the unknown parameters and the probability
distribution $\mathbf{p}=\{p(j,l)\dvtx j=0,\ldots,J;l=0,\ldots,L\}$ of the
covariates.
\begin{teo}\label{thmoptpi}
For every chosen inferential criterion \emph{C1--C5}, at
each stratum $(t_j,w_l)$ the optimal compound target $\pi_{\omega
}^*(j,l)$ satisfies the following properties:
\begin{itemize}
\item $\pi_{\omega}^*(j,l)\in(0,1)$ is a continuous function of
$\alpha
$, $\boldsymbol{\tau}$ and $\mathbf{p}$, that is, $\pi_{\omega
}^*(j,l)=\pi_{\omega}^*(j,l;\alpha,\boldsymbol{\tau},\mathbf
{p})$, and
it is increasing in $\theta(j,l)$;
\item if $(\alpha', \boldsymbol{\tau}')$ and $(\alpha'',
\boldsymbol
{\tau}'')$ are parameter values with corresponding ethical gains
$\theta
'(j,l)=-\theta''(j,l)$, then for any given covariate distribution
$\mathbf{p}$,
%
\begin{equation}\label{thmoptpi2}
\pi_{\omega}^*(j,l;\alpha',\boldsymbol{\tau}',\mathbf{p})=1-\pi
_{\omega
}^*(j,l;\alpha'',\boldsymbol{\tau}'',\mathbf{p}),
\end{equation}
so that the optimal compound target always assigns more than half the
subjects to the better treatment;\vadjust{\goodbreak}
\item if $\theta(j,l)>0$, then $\pi_{\omega}^*(j,l)$ is increasing in
$p(j,l)$, whereas when $\theta(j,l)<0$, then $\pi_{\omega}^*(j,l)$ is
decreasing in $p(j,l)$.
\end{itemize}
\end{teo}
\begin{pf}
For any given stratum $(t_j,w_l)$, let $k(\pi_n(j,l))=\partial
\widetilde{\Phi}_{I}(\boldsymbol{\pi}_n)/\break \partial\pi_n(j,l)$.
From the
convexity of the inferential criterion, the function $k(\cdot)$ is
monotonically increasing with $k(1/2)=0$, due to the optimality of the
jointly balanced design in (\ref{InfOpti}). Furthermore, the right-hand
side of (\ref{compound target}) is a~continuous function of $\alpha$,
$\boldsymbol{\tau}$ and $\mathbf{p}$, due to the properties of
$\omega(\cdot)$; it is straightforward to check that, for any given
$p(j,l)$, it is also an increasing function of~$\theta(j,l)$ and is
monotone in $p(j,l)$ (increasing when $\theta(j,l)>0$ and decreasing if
$\theta(j,l)<0$). Since the sign of the left-hand side of (\ref
{compound target}) depends only on the sign of $k(\cdot)$, if $\theta
(j,l)>0$ (i.e., $A$ is better than $B$ for this stratum), then the
right-hand side of (\ref{compound target}) is positive and thus the
optimal compound target $\pi_{\omega}^*(j,l)>1/2$; otherwise, if
$\theta
(j,l)<0$, then $\pi_{\omega}^*(j,l)<1/2$ and $\theta(j,l)=0$ if and
only if $\pi_{\omega}^*(j,l)=1/2$. Moreover, observe that $\lim
_{\zeta
\rightarrow0}k(\zeta)=- \infty$ and $\lim_{\zeta\rightarrow
1}k(\zeta
)=+ \infty$ and thus $\pi_{\omega}^*(j,l) \neq\{0,1\}$, since
$\widetilde{\Phi}_{E}(\cdot)$ is limited.
By taking the derivative of the left-hand side of (\ref{compound
target}) with respect to $\pi_n(j,l)$ we obtain
\[
\frac{\partial k(\pi_n(j,l))}{\partial\pi_n(j,l)}[\widetilde{\Phi
}_{E}(\boldsymbol{\pi}_n)]^2+2 \widetilde{\Phi}_{E}(\boldsymbol
{\pi
}_n)k(\pi_n(j,l)) \theta(j,l)p(j,l),
\]
where the first term is always positive, due to the convexity of the
inferential criterion. Furthermore, locally around $\pi_{\omega}^*(j,l)$
\[
k(\pi_{\omega}^*(j,l))\theta(j,l)p(j,l)>0,
\]
since if $\theta(j,l)>0$, then $\pi_{\omega}^*(j,l)>1/2$ and $k(\pi
_{\omega}^*(j,l))>0$ (and, analogously, when $\theta(j,l)<0$). Thus, as
a function of $\pi_n(j,l)$ the left-hand side of (\ref{compound
target}) is locally increasing around $\pi_{\omega}^*(j,l)$, so that
$\pi_{\omega}^*(j,l)$ is a continuous function of~$\alpha$,
$\boldsymbol
{\tau}$ and $\mathbf{p}$ and it is increasing in $\theta(j,l)$ and
$p(j,l)$, due to the property of the right-hand side of (\ref{compound
target}).
Concerning (\ref{thmoptpi2}), for a given covariate
distribution~$\mathbf{p}$ the ethical criterion in (\ref{expcriteth}) can be
regarded as a function of $\pi_n(j,l)$ and $\theta(j,l)$ by letting
\[
\sum_{j=0}^{J} \sum_{l=0}^{L}
|\theta(j,l)| p(j,l) \biggl\{\frac{1}{2}-\biggl[\frac{1}{2}-\pi_n(j,l)\biggr]\sgn
(\theta
(j,l))\biggr \} = \upsilon+g(\pi_n(j,l),\theta(j,l)),
\]
where
\[
g(\pi_n(j,l), \theta(j,l))=|\theta(j,l)| p(j,l) \biggl\{\frac
{1}{2}-\biggl[\frac
{1}{2}-\pi_n(j,l)\biggr]\sgn(\theta(j,l)) \biggr\},
\]
so the left-hand side of (\ref{compound target}) can be rewritten as
$[\upsilon+g(\pi_n(j,l), \theta(j,l))]\cdot k(\pi_n(j,l))$.
First of all note that, for every chosen inferential criterion C1--C5 the function
$ k(\cdot)$ is symmetric around the point $(1/2;0)$ since
$k(1/2+\epsilon)=-k(1/2-\epsilon) $ for any $\epsilon\in(0;1/2)$.
Moreover, $g(1/2+\epsilon, \theta(j,l))= g(1/2-\epsilon, -\theta
(j,l))$, so that the left-hand side of (\ref{compound target}) is also
symmetric around $(1/2;0)$, and this implies the symmetric property of
the compound target.
\end{pf}

\subsection{Example: the inferential criteria based on the
determinant}\label{example1}
$\!\!\!$From~(\ref{Dopt}) and (\ref{C3_crit}) it is easy to see that C1 and C2
 have the same standardized version,
%
\begin{equation}\label{D_optimality}
\Psi_{I}(\boldsymbol{\pi}_n)=4^{(J+1) (L+1)} \prod_{j=0}^{J} \prod
_{l=0}^{L} \pi_n(j,l) [1-\pi_n(j,l)].
\end{equation}
Assuming now two binary covariates and two different scenarios for
their population distribution, that is, a uniform one
$\mathcal{U}$ where each stratum is equally represented, that is,
$p(0,0)=p(1,0)=p(0,1)=p(1,1)=0.25$, and a~nonuniform distribution
$\mathcal{NU}$ with $p(0,0)=0.2$, $p(1,0)=0.3$, $p(0,1)=0.4$,
$p(1,1)=0.1$. Table \ref{Tabella 1} shows the derived optimal compound
targets in the case of four different ethical weights, namely the cdf's
of a $\chi^2_{(r)}$ with $r=1$,~$2$ and $\omega_s$ in (\ref{omegas})
with $s=1, 2$.

The optimal compound target always assigns the majority of subjects to
the better treatment. The ethical weight increases as $r$ and $s$
decrease, and therefore less emphasis is given to the inferential
precision and more attention to ethical demands, as expected.
Furthermore, since criteria C1 and C2 treat every
stratum in the same way, when $p(j,l)=p(\widetilde{j},\widetilde{l})$
and $\theta(j,l)=1-\theta(\widetilde{j},\widetilde{l})$, then $\pi
_\omega^*(j,l)=1-\pi_\omega^*(\widetilde{j},\widetilde{l})$.

\subsection{Example: the inferential criteria based on the trace}
Consider two binary covariates under the same settings of the previous
example. By taking the approximation $E_{\mathbf
{Z}}[N_n(j,l)^{-1}]=[np(j,l)]^{-1}$, Table \ref{Tabella 2} shows the
optimal compound targets when the inferential criterion C3 is
adopted, whereas Table \ref{Tabella 3} deals with criterion C4
(or, equivalently, C5).
%

Note that, since C3--C5 treat the strata in a
different way, even if $p(0,1)\,{=}\break p(1,1)\,{=}\,0.25$ and $\theta(0,1)\,{=}\,-\theta
(1,1)\,{=}\,-1$, that is, when $(\alpha,\boldsymbol{\tau}^t)\,{=}\,(-4,-1, 3,3)$,
then $\pi_\omega^*(0,1) \,{\neq}\,1\,{-}\,\pi_\omega^*(1,1)$.

%
\begin{sidewaystable}
\tabcolsep=0pt
\caption{Optimal compound targets adopting criteria \emph{C1--C2}}\label{Tabella 1}
\begin{tabular*}{\textwidth}{@{\extracolsep{\fill}}llcccccccc@{}}
\hline
& & \multicolumn{4}{c}{$\boldsymbol{(\alpha,\tau^t)=(1,1,1,1)}$} &
\multicolumn{4}{c@{}}{$\boldsymbol{(\alpha,\tau^t)=(-4,-1,3,3)}$}\\[-5pt]
& & \multicolumn{4}{c}{\hrulefill} &\multicolumn{4}{c@{}}{\hrulefill}\\
& & $\boldsymbol{\theta(0,0)=1}$ & $\boldsymbol{\theta(1,0)=2}$ & $\boldsymbol{\theta(0,1)=2}$ & $\boldsymbol{\theta
(1,1)=4}$ & $\boldsymbol{\theta(0,0)=-4}$ & $\boldsymbol{\theta(1,0)=-5}$ &$\boldsymbol{\theta(0,1)=-1}$ &
$\boldsymbol{\theta(1,1)=1}$\\
& & $\boldsymbol{\pi_\omega^*(0,0)}$ & $\boldsymbol{\pi_\omega^*(1,0)}$ & $\boldsymbol{\pi_\omega
^*(0,1)}$ &
$\boldsymbol{\pi_\omega^*(1,1)}$ & $\boldsymbol{\pi_\omega^*(0,0)}$& $\boldsymbol{\pi_\omega^*(1,0)}$
&$\boldsymbol{\pi_\omega^*(0,1)}$& $\boldsymbol{\pi_\omega^*(1,1)}$\\
\hline
$\chi^2_{(1)}$ & $\mathcal{NU}$ & 0.578 & 0.700 &
0.743 & 0.646 & 0.278 & 0.186 & 0.371 & 0.534 \\
& $\mathcal{U}$ & 0.593 & 0.670 & 0.670 & 0.771 & 0.242 & 0.209 &
0.415 & 0.585 \\
{$\chi^2_{(2)}$} & $\mathcal{NU}$ & 0.544 & 0.623 &
0.660 & 0.587 & 0.352 & 0.264 & 0.421 & 0.520 \\
& $\mathcal{U}$ & 0.554 & 0.605 & 0.605 & 0.689 & 0.319 & 0.287 &
0.449 & 0.551 \\ [6pt]
{$\omega_{1}$} & $\mathcal{NU}$ & 0.537 & 0.606 & 0.637
& 0.572 & 0.353 & 0.265 & 0.421 & 0.520 \\
& $\mathcal{U}$ & 0.549 & 0.596 & 0.596 & 0.674 & 0.321 & 0.289 &
0.449 & 0.551 \\
{$\omega_2$} & $\mathcal{NU}$ & 0.521 & 0.562 & 0.581 &
0.541 & 0.397 & 0.324 & 0.447 & 0.513 \\
& $\mathcal{U}$ & 0.530 & 0.559 & 0.559 & 0.614 & 0.373 & 0.346 &
0.466 & 0.534 \\
\hline
\end{tabular*}\vspace*{10pt}
\tabcolsep=0pt
\caption{Optimal compound targets adopting criterion \emph{C3}}
\label{Tabella 2}
\begin{tabular*}{\textwidth}{@{\extracolsep{\fill}}llcccccccc@{}}
\hline
& & \multicolumn{4}{c}{$\boldsymbol{(\alpha,\tau^t)=(1,1,1,1)}$} &
\multicolumn{4}{c@{}}{$\boldsymbol{(\alpha,\tau^t)=(-4,-1,3,3)}$}\\[-5pt]
& & \multicolumn{4}{c}{\hrulefill} &\multicolumn{4}{c@{}}{\hrulefill}\\
& & $\boldsymbol{\theta(0,0)=1}$ & $\boldsymbol{\theta(1,0)=2}$ & $\boldsymbol{\theta(0,1)=2}$ & $\boldsymbol{\theta
(1,1)=4}$ & $\boldsymbol{\theta(0,0)=-4}$ & $\boldsymbol{\theta(1,0)=-5}$ &$\boldsymbol{\theta(0,1)=-1}$ &
$\boldsymbol{\theta(1,1)=1}$\\
& & $\boldsymbol{\pi_\omega^*(0,0)}$ & $\boldsymbol{\pi_\omega^*(1,0)}$ & $\boldsymbol{\pi_\omega
^*(0,1)}$ &
$\boldsymbol{\pi_\omega^*(1,1)}$ & $\boldsymbol{\pi_\omega^*(0,0)}$& $\boldsymbol{\pi_\omega^*(1,0)}$
&$\boldsymbol{\pi_\omega^*(0,1)}$& $\boldsymbol{\pi_\omega^*(1,1)}$\\
\hline
{$\chi^2_{(1)}$} & $\mathcal{NU}$ & 0.658 & 0.868 &
0.900 & 0.805 & 0.179 & 0.077 & 0.128 & 0.677 \\
& $\mathcal{U}$ & 0.697 & 0.835 & 0.835 & 0.916 & 0.154 & 0.099 &
0.214 & 0.846 \\
{$\chi^2_{(2)}$} & $\mathcal{NU}$ & 0.572 & 0.792 &
0.841 & 0.706 & 0.277 & 0.125 & 0.205 & 0.582 \\
& $\mathcal{U}$ & 0.598 & 0.745 & 0.745 & 0.866 & 0.241 & 0.158 &
0.318 & 0.759 \\[6pt]
{$\omega_1$} & $\mathcal{NU}$ & 0.557 & 0.767 & 0.821 &
0.678 & 0.279 & 0.126 & 0.206 & 0.581 \\
& $\mathcal{U}$ & 0.586 & 0.728 & 0.728 & 0.856 & 0.243 & 0.159 &
0.320 & 0.757 \\
{$\omega_2$} & $\mathcal{NU}$ & 0.530 & 0.696 & 0.760 &
0.610 & 0.346 & 0.169 & 0.268 & 0.546 \\
& $\mathcal{U}$ & 0.548 & 0.658 & 0.658 & 0.806 & 0.308 & 0.210 &
0.382 & 0.692 \\
\hline
\end{tabular*}\vspace*{-15pt}
\end{sidewaystable}
%
\begin{sidewaystable}
\tabcolsep=0pt
\caption{Optimal compound targets adopting criteria \emph{C4} or
\emph{C5}}\label{Tabella 3}
\begin{tabular*}{\textwidth}{@{\extracolsep{\fill}}llcccccccc@{}}
\hline
& & \multicolumn{4}{c}{$\boldsymbol{(\alpha,\tau^t)=(1,1,1,1)}$} &
\multicolumn{4}{c@{}}{$\boldsymbol{(\alpha,\tau^t)=(-4,-1,3,3)}$}\\[-5pt]
& & \multicolumn{4}{c}{\hrulefill} &\multicolumn{4}{c@{}}{\hrulefill}\\
& & $\boldsymbol{\theta(0,0)=1}$ & $\boldsymbol{\theta(1,0)=2}$ & $\boldsymbol{\theta(0,1)=2}$ & $\boldsymbol{\theta
(1,1)=4}$ & $\boldsymbol{\theta(0,0)=-4}$ & $\boldsymbol{\theta(1,0)=-5}$ &$\boldsymbol{\theta(0,1)=-1}$ &
$\boldsymbol{\theta(1,1)=1}$\\
& & $\boldsymbol{\pi_\omega^*(0,0)}$ & $\boldsymbol{\pi_\omega^*(1,0)}$ & $\boldsymbol{\pi_\omega
^*(0,1)}$ &
$\boldsymbol{\pi_\omega^*(1,1)}$ & $\boldsymbol{\pi_\omega^*(0,0)}$& $\boldsymbol{\pi_\omega^*(1,0)}$
&$\boldsymbol{\pi_\omega^*(0,1)}$& $\boldsymbol{\pi_\omega^*(1,1)}$\\
\hline
{$\chi^2_{(1)}$} & $\mathcal{NU}$ & 0.677 & 0.860 &
0.895 & 0.795 & 0.166 & 0.082 & 0.137 & 0.663 \\
& $\mathcal{U}$ & 0.717 & 0.827 & 0.827 & 0.912 & 0.142 & 0.105 &
0.225 & 0.837 \\
{$\chi^2_{(2)}$} & $\mathcal{NU}$ & 0.585 & 0.782 &
0.833 & 0.694 & 0.259 & 0.133 & 0.217 & 0.573 \\
& $\mathcal{U}$ & 0.615 & 0.734 & 0.734 & 0.859 & 0.223 & 0.167 &
0.331 & 0.747 \\[6pt]
{$\omega_{1}$} & $\mathcal{NU}$ & 0.567 & 0.756 & 0.812
& 0.666 & 0.261 & 0.134 & 0.218 & 0.572 \\
& $\mathcal{U}$ & 0.601 & 0.717 & 0.717 & 0.849 & 0.225 & 0.169 &
0.333 & 0.744 \\
{$\omega_2$} & $\mathcal{NU}$ & 0.536 & 0.685 & 0.749 &
0.601 & 0.328 & 0.179 & 0.282 & 0.541 \\
& $\mathcal{U}$ & 0.558 & 0.645 & 0.645 & 0.797 & 0.289 & 0.221 &
0.393 & 0.679 \\
\hline
\end{tabular*}
\end{sidewaystable}

In general, when $r=s$ the ethical skew is larger if we adopt the cdf
of~$\chi^2_{(r)}$ w.r.t. $\omega_s$, and this skew is particularly high
for $r=1$, which could induce strong imbalances among the treatment
groups (see Tables \ref{Tabella 1}--\ref{Tabella 3} with $(\alpha
,\boldsymbol{\tau}^t)=(-4,-1,3,3)$) and, consequently, a great loss of
inferential efficiency. This behavior suggests that the adoption of the
cdf of $\chi^2_{(1)}$ is adequate only in situations with prohibitive
ethical costs.

%
\begin{table}
\tabcolsep=0pt
\caption{Constrained optimal targets adopting criteria \emph
{C1--C2} and Bandyopadhyay and Biswas's allocations $\pi_{T}$,
with $(\alpha,\boldsymbol{\tau}^t)=(1,1,1,1)$ and uniform
distribution}\label{Tabella 35}
\begin{tabular*}{\textwidth}{@{\extracolsep{\fill}}lcccccc@{}}
\hline
& $\boldsymbol{\theta(0,0)=1}$ & $\boldsymbol{\theta(1,0)=2}$ & $\boldsymbol{\theta(0,1)=2}$ & $\boldsymbol{\theta
(1,1)=4}$& $\boldsymbol{\Psi_E}$ & $\boldsymbol{\Psi_I=C}$\\
\hline
& $\pi_{\omega_C}^*(0,0)$ & $\pi_{\omega_C}^*(1,0)$ & $\pi_{\omega
_C}^*(0,1)$ & $\pi_{\omega_C}^*(1,1)$ & & \\
$\omega_C=0.356$ & 0.523 & 0.546 & 0.546 & 0.589 & 0.56 & \hphantom{1}0.95 \\
$\omega_C=0.483$ & 0.528 & 0.566 & 0.566 & 0.612 & 0.59 & 0.9 \\
$\omega_C=0.700 $ &0.558 & 0.612 & 0.612 & 0.698 & 0.64 & \hphantom{1}0.75 \\
$\omega_C=0.883$ & 0.599 & 0.679 & 0.679 & 0.781 & 0.72 & 0.5 \\
$\omega_C=0.969 $ & 0.656 & 0.756 & 0.756 & 0.851 & 0.79 & \hphantom{1}0.25 \\
[6pt]
& $\pi_{T}(0,0)$ & $\pi_{T}(1,0)$ & $\pi_{T}(0,1)$ & $\pi_{T}(1,1)$ & & \\
$T=1$ & 0.841 & 0.977 & 0.977 & 0.999 & 0.97 & $10^{-6}$ \\
$T=2$ & 0.691 & 0.841 & 0.841 & 0.977 & 0.88 & \hphantom{1}0.02 \\
$T=3$ & 0.631 & 0.748 & 0.748 & 0.909 & 0.81 & \hphantom{1}0.17 \\
\hline
\end{tabular*}
\end{table}

\begin{remark}\label{constrained}
Using the analytical expressions of the inferential criteria given in
Proposition \ref{mainprop}, by (\ref{PsiE}) it is also possible to
derive optimal targets via a\vadjust{\goodbreak} constrained optimization approach. In the
same spirit of~\cite{Tym07}, the problem lies in finding the allocation
that maximizes the ethical impact for a chosen inferential efficiency.
In our context this corresponds to minimize~$\Psi_{E}^{-1}$ under the
constraint $\Psi_{I}^{-1}\leq C^{-1}$ for a prefixed constant $C < 1$
(we exclude the degenerate case $C=1$ that corresponds to assume $\Psi
_I=1$, i.e., no ethical concerns), representing a special case of our
combined optimization approach; see also \cite{Bag2010,CLY96,COOK94}.
Indeed,\vspace*{1pt} due to the strict convexity of both $\Psi_{E}^{-1}$ and $\Psi
_{I}^{-1}$, this is a convex optimization problem, and therefore the
Karush--Kuhn--Tucker (KKT) first order conditions are necessary and
sufficient and guarantee a unique optimal solution $\boldsymbol{\pi}
^*$ solving
%
\begin{equation}\label{KKT}
\nabla(\Psi_{E}^{-1}(\boldsymbol{\pi}^*))+\kappa\nabla(\Psi
_{I}^{-1}(\boldsymbol{\pi}^*))=\mathbf{0},
\end{equation}
where $\kappa\geq0$ is the KKT multiplier and
%
\begin{equation}\label{KKT1}
\kappa\bigl( \Psi_{I}^{-1}(\boldsymbol{\pi}^*) - C^{-1}\bigr)=0.
\end{equation}
First, note that (\ref{KKT}) corresponds to (\ref{compound target})
with $\kappa=(1-\omega)/\omega$ (where clearly $\omega\neq0$, since
$C \neq1$); thus, the candidate solution $\boldsymbol{\pi}^*$ belongs
to the class of compound optimal targets and should satisfy $\Psi
_{I}^{-1}(\boldsymbol{\pi}^*) = C^{-1}$, since $\kappa> 0$. By using
the same arguments of the proof of Theorem \ref{thmoptpi}, any compound
target $\boldsymbol{\pi}_{\omega}^*$ solving (\ref{compound
target}) is
a continuous and monotone function of $\omega$ and, from the properties
of $\Psi_{I}^{-1}$, for any fixed $C < 1$ there exists a unique
constant weight $\omega=\omega_C$ such that
$\boldsymbol{\pi}^*=\boldsymbol{\pi}_{\omega_C}^*$ satisfies (\ref
{KKT1}).

As a numerical example, consider now the standardized inferential
criterion (\ref{D_optimality})
in the case of two binary covariates with $(\alpha,\boldsymbol{\tau
}^t)=(1,1,1,1)$ and uniform distribution $\mathcal{U}$. The upper block
of Table \ref{Tabella 35} shows the derived constrained optimal targets
as $C$ varies, together with the corresponding ethical weight $\omega
_C$ of our combined optimization approach and the value of the ethical
criterion $\Psi_E$. Moreover, the bottom part of the Table gives the
asymptotic\vadjust{\goodbreak} allocations of the procedure suggested by Bandyopadhyay and
Biswas \cite{Ban01} (and further analyzed in \cite{Atk05a}), which are
given at each stratum $(j,l)$ by the standard normal cdf evaluated at
$\theta(j,l)/T$, for $T=1, 2$ and $3$.
For instance, an inferential efficiency equal to 75\% under the
constrained approach corresponds to an ethical weight equal to 70\% in
the combined framework; whereas, assuming $C=0.25$ means that the role
of ethics is almost dominant in the combined optimization. Clearly, the
arbitrary choice of the constant $C$ in the constrained setting can be
directly translated in the subjective choice of $\omega$ in the
combined approach; however, the possibility of modeling the ethical
weight as a function of the unknown parameters allows us to
discriminate among different situations that could be a-priori only
partially known or, more commonly, completely unknown. For instance, if
we set $\omega_C=0.483$, then the 38.8\% of subjects within the stratum
$(1,1)$ will receive the worst treatment; whereas, the percentage of
allocations to the worst treatment is only 22.9\% or 32,6\% if the cdf
of $\chi^2_{(1)}$ or $\omega_1$ are assumed as weight functions,
respectively; see Table \ref{Tabella 1}. Moreover, the allocations
proposed by Bandyopadhyay and Biswas (2001) are strongly skewed toward
the better treatment, so that the inferential precision collapses. This
is particularly true for small values of $T$, as emphasized in \cite{Ros08},
 and this behavior is also confirmed, even if we adopt criteria
C3--C5.
\end{remark}

\section{The reinforced doubly adaptive biased coin design}\label{s7}
As shown previously, the compound target allocation $\boldsymbol{\pi
}_{\omega}^*$ depends in general on the unknown parameters of
the model and, since this function is continuous, covariate-adjusted
response-adaptive procedures may be called for. These designs use at
each step the observed responses, the covariates and the previous
assignments, as well as the covariate profile of the current subject,
to modify the allocations as the experiment goes along in order to
gradually approach the desired target.

In this section we introduce the reinforced doubly adaptive biased coin
design. This is a general class of CARA procedures, which admits both
continuous and discontinuous randomization functions, aimed at
targeting any chosen allocation proportion by forcing closeness to the
target when necessary.

Let now $\boldsymbol{\pi}^{\ast}=\{\pi^{\ast}(j,l)\dvtx j=0,\ldots
,J;l=0,\ldots,L\}$ be a desired allocation such that, at each stratum
$(t_j,w_l)$, $\pi^{\ast}(j,l)\in(0,1)$ is a continuous function of the
unknown model parameters $\boldsymbol{\gamma}$ and $\mathbf{p}$.
Suppose that patients come to the trial sequentially and are assigned
to either treatment. Starting with $m$ observations on each treatment,
usually assigned by
using restricted randomization, an initial
non-trivial parameter estimation $\widehat{\boldsymbol{\gamma}}_{2m}$
and $\widehat{\mathbf{p}}_{2m}$ is derived. Then, at
each step $n$ ($n>2m)$ let $\widehat{\boldsymbol{\gamma}}_{n}$ and
$\widehat{\mathbf{p}}_{n}$
be the estimators of the parameters based on the first $n$
observations, where we assume them consistent in case of i.i.d.
observations, so that the
optimal target will be estimated by all the\vadjust{\goodbreak} data up to that step by
$\widehat{\boldsymbol{\pi}}_{n}^{\ast}=\boldsymbol{\pi}^{\ast
}(\widehat{\boldsymbol{\gamma}}_{n},\widehat{\mathbf{p}}_{n})$.
When the $(n+1)$st patient
with covariate profile $Z_{n+1}=(t_j,w_l)$ is ready to be randomized,
the reinforced doubly adaptive biased coin design assigns to him/her
treatment~$A$ with probability
%
\begin{equation}\label{phi}
\varphi(\pi_{n}(j,l);\widehat{\pi}_{ n}^{\ast}(j,l);\widehat{p}_{n}(j,l)),
\end{equation}
where the function $\varphi(x,y,z)\dvtx(0,1)^3 \rightarrow[0,1]$ satisfies
the following \mbox{properties}:

\begin{longlist}[(iii)]
\item[(i)] $\varphi$ is decreasing in $x$ and increasing in $y$, for
any $ z \in(0,1)$;
\item[(ii)] $\varphi(x,x,z)=x$ for any $z \in(0,1)$;
\item[(iii)] $\varphi$ is decreasing in $z$ if $x<y$, and increasing in
$z$ if $x>y$;
\item[(iv)] $\varphi(x,y,z)=1-\varphi(1-x,1-y,z)$ for any $ z \in(0,1)$.
\end{longlist}
Adopting the RDBCD in (\ref{phi}), within each stratum the allocation
proportion will be forced to the corresponding target, since from
conditions (i)--(ii) when $x \geq y$, then $\varphi(x,y,z) \leq y$ and
if $x < y$, then $\varphi(x,y,z) > y$ for any $z \in(0,1)$.
Furthermore, condition (iv) simply guarantees that $A$ and $B$ are
treated symmetrically, whereas (iii) means that the allocation is
forced towards optimality increasingly as the representativeness
of the strata in the population decreases. This property is of great
importance since the convergence of the allocation proportion depends
on the number of subjects belonging to each stratum, and therefore it
is related to the population distribution of the covariates. This may
be particularly critical for small samples,
where some profiles could be strongly under-represented
so that, both from the ethical and inferential viewpoint, the
need to force the closeness to the target could be
greater.
\begin{remark}\label{per referee rincoglioniti}
In general, at each step the allocation probability (\ref{phi}) does
not depend only on the estimates of the unknown parameters related to
the stratum where the current subject belongs; in fact, $\widehat{\pi
}_{ n}^{\ast}(j,l)$ could involve the estimation of the entire set of
parameters $\boldsymbol{\gamma}$ and $\mathbf{p}$. For example, if we
adopt the optimal compound target $\boldsymbol{\pi}_{\omega}^*$, at
each stratum $\pi_{\omega}^*(j,l)$ depends on~$\alpha$, $\boldsymbol
{\tau}$ and $\mathbf{p}$ [see, for instance, (\ref{compound target})],
so that (\ref{phi}) depends on the information gathered up to that step
from all the strata.
\end{remark}

Observe that we do not assume the continuity of $\varphi$, and
therefore it is possible to consider discontinuous randomization
functions. For instance, a~natural extension of the ERADE proposed by
\cite{Hu09} in the presence of covariates is
%
\begin{equation}\label{ERADE}
\varphi_{\mathrm{ERADE}} (x;y;z)=
\cases{1- \rho(1-y), & \quad $x<y$, \cr
y, & \quad $x=y$,\qquad $\forall z\in(0,1)$, \cr
\rho y, & \quad $x>y$}
\end{equation}
where the constant $\rho\in[0,1)$ controls the degree of randomness.
\begin{remark}\label{alldisco}
The randomization function in (\ref{phi}) could also be chosen in a
different way for each stratum $(t_j,w_l)$ by letting $\varphi=\varphi
_{jl}$, in order\vadjust{\goodbreak} to discriminate the importance of each of them and the
corresponding closeness to the target. Furthermore, the RDBCD can be
naturally extended to the case of several treatments.
\end{remark}

An interesting family of allocation functions belonging to the RDBCD is
%
\begin{eqnarray}\label{half_normal1}
\varphi(x;y;z)&=&{F\bigl[ D(x;y)^{H(z)} F^{-1}(y)\bigr]}\nonumber\\
&&{}/\bigl(F\bigl[ D(x;y)^{H(z)}
F^{-1}(y)\bigr]\\
&&\hphantom{{}/\bigl(}{}+ F\bigl[ D(1-x;1-y)^{H(z)} F^{-1}(1-y)\bigr]\bigr),\nonumber
\end{eqnarray}
where $F\dvtx\mathbb{R}^+\rightarrow\mathbb{R}^+$ is continuous and
strictly increasing, $H(z)$ is decreasing, while $D(x;y)\dvtx(0;1)^2
\rightarrow\mathbb{R}^+$ represents a dissimilarity measure between
the actual allocation proportion $x$ and the current estimate of the
optimal target~$y$, and $D$ is assumed to be decreasing in $x$ and
increasing in~$y$, with \mbox{$D(x;x)=1$}.
\begin{ex}
Letting $D(x;y)=1$ for any $(x,y)\in(0,1)^2$, then (\ref
{half_normal1}) corresponds to the CARA design analyzed by Zhang et al.
\cite{Zha07}, namely
%
\begin{equation}\label{gfunc}
\varphi_{Z}(x;y;z)=y \qquad \forall(x,z) \in(0,1)^2,
\end{equation}
which represents an analog of the sequential maximum likelihood design
in the presence of covariates.
Whereas if we let $F(t)=t$, $D(x;y)=y/x$ and $H(z)=\nu\geq0$ $\forall
z\in(0,1)$,
we obtain a natural extension in the presence of covariates of the
family of doubly adaptive biased coin designs, that is,
\[
\frac{y(y/x)^{\nu}}{y(y/x)^{\nu}+(1-y)[(1-y)/(1-x)]^{\nu}} \qquad \forall z
\in(0,1).
\]
However, note that the previous allocation function does not correspond
to the covariate-adjusted doubly adaptive biased coin design suggested
by Zhang and Hu in \cite{Zha09}, due to the fact that these authors
assume at each step $n$ a dissimilarity measure between the actual
allocation proportion $\pi_{n}(j,l)$ and the mean (over the steps) of
the estimates of the optimal target $n^{-1}\sum_{i=1}^{n}\widehat{\pi
}_{ i}^{\ast}(j,l)$, instead of the current estimate of the target
$\widehat{\pi}_{ n}^{\ast}(j,l)$ itself.
\end{ex}
\begin{ex}
If we set $F(t)=t^k$ with $k>0$, $D(x;y)=1-(x-y)$ and $H(z)=z^{-1}\
\forall z \in(0,1)$, then (\ref{half_normal1}) becomes
%
\begin{equation}\label{nostraphi1}\quad
\varphi_{BAZ_1} (x;y;z)=\frac{y[1-(x-y)]^{
{k}/{z}}}{y[1-(x-y)]^{{k}/{z}}+(1-y)[1-(y-x)]^{{k}/{z}}}.
\end{equation}
In order to account for discontinuous allocation functions, let for
instance $F(t)=t$, $H(z)=\{(J+1)(L+1)z\}^{-1}$ and
\[
D(x;y)= \cases{1+\varepsilon, &\quad $x<y$, \cr
1, & \quad $x=y$,\cr
1-\varepsilon, &\quad $x>y$,}\vadjust{\goodbreak}
\]
with $\varepsilon\in[0,1)$. Then (\ref{half_normal1}) becomes
%
\begin{eqnarray}\label{nostraphi2}
&&\varphi_{BAZ_2} (x;y;z)\nonumber
\\[-8pt]
\\[-8pt]
&& \qquad = \cases{\displaystyle\frac{y(1+\varepsilon)^{\{(J+1)(L+1)z\}^{-1}}}{y(1+\varepsilon)^{\{
(J+1)(L+1)z\}^{-1}} +(1-y)(1-\varepsilon)^{\{(J+1)(L+1)z\}^{-1}}} ,\vspace*{2pt}\cr  \hphantom{y,  \qquad}\, x<y, \cr
y,  \qquad x=y,\cr
\displaystyle\frac{y(1-\varepsilon)^{\{(J+1)(L+1)z\}^{-1}}}{y(1-\varepsilon)^{\{
(J+1)(L+1)z\}^{-1}}+(1-y)(1+\varepsilon)^{\{(J+1)(L+1)z\}^{-1}}} ,\vspace*{2pt}\cr  \hphantom{y,  \qquad}\,
x>y,}\nonumber
\end{eqnarray}
which allows us to force the allocations toward the chosen target
increasingly the more we move away from the uniform
distribution, maintaining at the same time a good degree of randomness.
\end{ex}

\begin{remark}\label{alldisco}
If we assume only an inferential viewpoint by letting \mbox{$\omega=0$}, then
the optimal target is the jointly balanced allocation in (\ref
{InfOpti}), so that the allocation probability (\ref{phi}) corresponds
to a stratified randomization. For instance, letting $F(t)=t^2$,
$H(z)=1$ and $D(x;y=1/2)=1-2(x-1/2)$, then procedure (\ref
{half_normal1}) corresponds to the $D_A$-optimal design proposed by
Atkinson \cite{Atk82},
\[
\frac{[1-\pi_{n}(j,l)]^2}{[1-\pi_{n}(j,l)]^2+\pi_{n}(j,l)^2};
\]
see the supplementary data in \cite{Baz11} for details.
\end{remark}

Let $\boldsymbol{\pi}^*_{t_j \otimes\mathbb{W}}=(\pi^*(j,1)p(j,1),
\ldots, \pi^*(j,L)p(j,L))^t$ and $\widetilde{{\boldsymbol{\pi}}}^{*
t}=(\boldsymbol{\pi}^{*t} _{\mathbb{T}},\boldsymbol{\pi}^{*
t}_{\mathbb
{W}},\break \boldsymbol{\pi}^{* t}_{\mathbb{T} \otimes\mathbb{W}})$, where
\begin{eqnarray*}
\boldsymbol{\pi}^*_{\mathbb{T}}&=&\Biggl(\sum_{l=0}^L\pi^*(1,l)p(1,l),
\ldots,\sum_{l=0}^L\pi^*(J,l)p(J,l)\Biggr)^t,\\
\boldsymbol{\pi}^*_{\mathbb{W}}&=&\Biggl(\sum_{j=0}^J\pi^*(j,1)p(j,1),
\ldots,\sum_{j=0}^J\pi^*(j,L)p(j,L)\Biggr)^t
\end{eqnarray*}
and $\boldsymbol{\pi}^*_{\mathbb{T} \otimes\mathbb{W}}$ given by
\begin{eqnarray*}
&&(\pi^*(1,1)p(1,1), \ldots, \pi^*(1,L)p(1,L), \ldots,\\
&&\qquad \pi^*(J,1)p(J,1),
\ldots,\pi^*(J,L)p(J,L))^t.
\end{eqnarray*}
The following theorem establishes the strong consistency of both the
allocation proportion and the estimator of the target, as well as the
strong consistency and asymptotic normality of the estimators of the
unknown parameters of the model.
\begin{teo}\label{thconvergence}
$\!\!\!$For any given target allocation $\boldsymbol{\pi}^{\ast}\in
(0,1)^{(J+1)\cdot(L+1)}$ which is a continuous function of the unknown
model parameters $\boldsymbol{\gamma}$ and $\mathbf{p}$, then adopting
the reinforced doubly adaptive biased coin design (\ref{phi}), as $n$
tends to infinity
%
\begin{equation}\label{Istat}
\boldsymbol{\pi}_n \rightarrow\boldsymbol{\pi}^* \qquad \mbox{a.s.}
\quad \mbox{and}\quad
 \boldsymbol{\widehat{\pi}}_{n }^{\ast}\rightarrow\boldsymbol
{\pi}^* \qquad \mbox{a.s.}
\end{equation}
Furthermore,
%
\begin{equation}\label{Istat1}
\qquad\boldsymbol{\hat\gamma}_n \rightarrow\boldsymbol{\gamma} \qquad \mbox{a.s.}
\quad \mbox{and} \quad \sqrt{n}(\boldsymbol{\hat\gamma}_n-\boldsymbol{\gamma
})\rightarrow
N(\mathbf{0};\boldsymbol{\mathcal{M}}^{-1}(\boldsymbol{\pi}^{\ast
})) \qquad \mbox{in law},
\end{equation}
where $\boldsymbol{\mathcal{M}}(\boldsymbol{\pi}^{\ast})$ is
\[  \left(
\begin{array} {c@{\quad}c@{\hspace*{6pt}}|@{\hspace*{6pt}}c@{\quad}c}
\displaystyle\sum_{j=0}^{J}\sum_{l=0}^{L}\pi^*(j,l)p(j,l) & 0 & \widetilde
{{\boldsymbol{\pi}}}^{* t} & \mathbf{0}_{1\times p}\\
0 & \displaystyle 1-\sum_{j=0}^{J}\sum_{l=0}^{L} \pi^*(j,l)p(j,l) &\mathbf
{0}_{1\times p} & \mathbf{1}_{p}^t-\widetilde{{\boldsymbol{\pi
}}}^{* t}\\
[-6.6pt]\\[-6pt] \hline\vspace*{-6pt}\\[-7pt]
\widetilde{{\boldsymbol{\pi}}}^{*} & \mathbf{0}_{p\times1}&
\boldsymbol{\mathcal{M}}^A & \mathbf{0}_{p\times p}\\
\mathbf{0}_{p\times1} & \mathbf{1}_p-\widetilde{{\boldsymbol{\pi
}}}^{*} & \mathbf{0}_{p\times p} & \boldsymbol{\mathcal{M}}^B\\
\end{array}
\right),
\]
with
%
\begin{equation}\label{M22A}
\boldsymbol{\mathcal{M}}^A=\left(
\begin{array} {c@{\hspace*{6pt}}|@{\hspace*{6pt}}c}
\boldsymbol{\mathcal{M}}^A_{11} & \boldsymbol{\mathcal{M}}^A_{12}\\
[-6.6pt]\\[-6pt] \hline\vspace*{-5pt}\\[-7pt]
(\boldsymbol{\mathcal{M}}^{A}_{12})^t & \boldsymbol{\mathcal
{M}}^A_{22}
\end{array}
\right),
\end{equation}
such that $\boldsymbol{\mathcal{M}}_{22}^A=\diag(\boldsymbol{\pi
}^*_{\mathbb{T} \otimes
\mathbb{W}})$,
\begin{eqnarray*}
\boldsymbol{\mathcal{M}}^A_{11}&=&\left(
\begin{array} {c@{\quad}c@{\quad}c@{\hspace*{6pt}}|@{\hspace*{6pt}}c}
&&& \boldsymbol{\pi}^{* t}_{t_1 \otimes\mathbb{W}}\\
&\diag(\boldsymbol{\pi}^*_{\mathbb{T}})&& \vdots\\
&&& \boldsymbol{\pi}^{* t}_{t_J \otimes\mathbb{W}} \\
[-6.6pt]\\[-6pt] \hline\vspace*{-5pt}\\[-7.6pt]
\boldsymbol{\pi}^*_{t_1 \otimes\mathbb{W}}& \ldots& \boldsymbol
{\pi
}^*_{t_J \otimes\mathbb{W}} & \diag(\boldsymbol{\pi}^*_{\mathbb{W}})
\end{array}
\right),\\
\boldsymbol{\mathcal{M}}^A_{12}&=&\left(
\begin{array}{c@{\quad}c@{\quad}c@{\quad}c@{\quad}c}
\boldsymbol{\pi}^{* t}_{t_1 \otimes\mathbb{W}} & \mathbf{0}&
\mathbf{0} & \ldots& \mathbf{0}\\
\mathbf{0} &\boldsymbol{\pi}^{* t}_{t_2 \otimes\mathbb{W}} &
\mathbf{0} & \ldots& \mathbf{0}\\
\vdots& & \ddots& & \mathbf{0}\\
\mathbf{0} & \mathbf{0} & \ldots& & \boldsymbol{\pi}^{* t}_{t_J
\otimes\mathbb{W}}\\
[-6.6pt]\\[-6pt] \hline\vspace*{-5pt}\\[-7.6pt]
\diag(\boldsymbol{\pi}^*_{t_1 \otimes\mathbb{W}}) & \diag
(\boldsymbol{\pi}^*_{t_2 \otimes\mathbb{W}}) & \ldots& & \diag(\boldsymbol
{\pi}^*_{t_J \otimes\mathbb{W}})\\
\end{array}
\right).
\end{eqnarray*}
Moreover, $\boldsymbol{\mathcal{M}}^B$ is partitioned similarly to
(\ref
{M22A}) with treatment $A$ replaced by $B$, that is, $\boldsymbol
{\mathcal{M}}_{22}^B=\diag(\mathbf{1}_{JL}-\boldsymbol{\pi
}^*_{\mathbb
{T} \otimes\mathbb{W}})$,
{\fontsize{10.6}{12.6}\selectfont{\begin{eqnarray*}
\boldsymbol{\mathcal{M}}_{11}^B&=&\left(
\begin{array} {@{}c@{\ \ }c@{\ \ }c@{\ \ }|@{\ \ }c@{}}
&&& \mathbf{1}_L^t-\boldsymbol{\pi}^{* t}_{t_1 \otimes\mathbb{W}}\\
&\diag(\mathbf{1}_J-\boldsymbol{\pi}^*_{\mathbb{T}})& & \vdots\\
&& & \mathbf{1}_L^t-\boldsymbol{\pi}^{* t}_{t_J \otimes\mathbb{W}}
\\
[-6.6pt]\\[-6pt] \hline\vspace*{-5pt}\\[-7.6pt]
\mathbf{1}_L-\boldsymbol{\pi}^*_{t_1 \otimes\mathbb{W}}& \ldots&
\mathbf{1}_L-\boldsymbol{\pi}^*_{t_J \otimes\mathbb{W}} & \diag
(\mathbf{1}_L-\boldsymbol{\pi}^*_{\mathbb{W}})
\end{array}
\right),\\
\boldsymbol{\mathcal{M}}_{12}^B&=&\left(
\begin{array}{@{}c@{\ \ }c@{\ \ }c@{\ \ }c@{\ \ }c@{}}
\mathbf{1}_L^t-\boldsymbol{\pi}^{* t}_{t_1 \otimes\mathbb{W}} &
\mathbf
{0}& \mathbf{0} & \ldots& \mathbf{0}\\
\mathbf{0} &\mathbf{1}_L^t-\boldsymbol{\pi}^{* t}_{t_2 \otimes
\mathbb
{W}} & \mathbf{0} & \ldots& \mathbf{0}\\
\vdots& & \ddots& & \mathbf{0}\\
\mathbf{0} & \mathbf{0} & \ldots& & \mathbf{1}_L^t- \boldsymbol
{\pi
}^{* t}_{t_J \otimes\mathbb{W}}\\
[-6.6pt]\\[-6pt] \hline\vspace*{-6pt}\\[-7pt]
\diag(\mathbf{1}_L-\boldsymbol{\pi}^*_{t_1 \otimes\mathbb{W}}) &
\diag(\mathbf{1}_L-\boldsymbol{\pi}^*_{t_2 \otimes\mathbb{W}}) &
\ldots& & \diag(\mathbf{1}_L-\boldsymbol{\pi}^*_{t_J \otimes
\mathbb
{W}})\\
\end{array}
\right)\!.
\end{eqnarray*}}}%
\end{teo}
\begin{pf}
See Appendix \ref{a2}.
\end{pf}

Note that the asymptotic normality of the allocation proportions can be
derived as in \cite{Hu04,Hu09,Zha07} by adding suitable conditions of
differentiability for the target $\boldsymbol{\pi}^*$ and, eventually,
for $\varphi$.
\begin{cor}\label{corconvergence}
Let the optimal compound target $\boldsymbol{\pi}_{\omega}^*$ be the
desired allocation, then adopting the RDBCD in (\ref{phi}), as $n$
tends to infinity
\begin{eqnarray*}
\boldsymbol{\pi}_n &\rightarrow& \boldsymbol{\pi}_{\omega}^* \qquad \mbox{a.s.}
\quad \mbox{and} \quad\boldsymbol{\widehat{\pi}}_{n }^{\ast}\rightarrow
\boldsymbol{\pi}_{\omega}^*  \qquad \mbox{a.s.} \\
\boldsymbol{\hat\gamma}_n &\rightarrow& \boldsymbol{\gamma} \qquad \mbox{a.s.}
\quad\mbox{and} \quad \sqrt{n}(\boldsymbol{\hat\gamma}_n-\boldsymbol{\gamma
})\rightarrow
N(\mathbf{0};\boldsymbol{\mathcal{M}}^{-1}(\boldsymbol{\pi
}_{\omega}^*)) \qquad \mbox{in law}.
\end{eqnarray*}
\end{cor}

\section{Finite sample properties}\label{s8}
In order to perform some finite sample comparisons, we have conducted a
simulation study by adopting the inferential criterion C1
and assuming as ethical weight function $\omega$ the cdf of $\chi
^2_{(1)}$. Moreover, we have taken into account normal responses with
$\sigma^2=1$ and two binary covariates with the previously used
settings, that is, (i) two population scenarios, namely the uniform
distribution $\mathcal{U}$ and the nonuniform one $\mathcal{NU}$ with
$p(0,0)=0.2$, $p(1,0)=0.3$, $p(0,1)=0.4$, $p(1,1)=0.1$, (ii) two
parameter settings: $(\alpha,\boldsymbol{\tau}^t)=(1,1,1,1)$ and
$(\alpha,\boldsymbol{\tau}^t)=(-4,-1,3,3)$. The results come from
$h=500$ simulations with $m=4$ and $n=500$. Concerning the continuous
randomization functions, we consider the CARA design $\varphi_{Z}$ in
(\ref{gfunc}) and the RDBCD in (\ref{nostraphi1}) with $k=1$. As
regards the discontinuous case, we perform the simulations adopting
$\varphi_{BAZ_2}$ in (\ref{nostraphi2}) and $\varphi_{\mathrm{ERADE}}$
in~(\ref{ERADE}), where we put $\varepsilon=\rho=2/3$, for homogeneity.
Expectation and standard deviation (within brackets) of the proportion
of allocations to treatment $A$ are given in Tables \ref{Tabella
4}--\ref{Tabella 7}.
%
\begin{table}
\def\arraystretch{0.9}
\tabcolsep=0pt
\caption{Expectation and standard deviation (within brackets) of the
proportion of allocations to $A$ under the uniform distribution
$\mathcal{U}$ with $(\alpha,\boldsymbol{\tau}^t)=(1,1,1,1)$, so that
the optimal compound target is $\pi_\omega^*(0,0)=0.593$, $\pi
_\omega
^*(1,0)=\pi_\omega^*(0,1)=0.670$ and $\pi_\omega^*(1,1)=0.771$}
\label{Tabella 4}
\begin{tabular*}{\textwidth}{@{\extracolsep{\fill}}lcccc@{}}
\hline
& $\boldsymbol{\pi_n(0,0)}$ & $\boldsymbol{\pi_n(1,0)}$ & $\boldsymbol{\pi_n(0,1)}$ & $\boldsymbol{\pi_n(1,1)}$ \\
\hline
{$\varphi_{Z}$} & 0.592 & 0.667 & 0.666 & 0.764 \\
&(0.051) & (0.049) & (0.045)& (0.041) \\
[3pt]
{$\varphi_{BAZ_1}$} & 0.592 & 0.667 & 0.670 & 0.768 \\
& (0.027) & (0.027) & (0.026) & (0.025) \\
[3pt]
{$\varphi_{BAZ_2}$} & 0.591 & 0.668 & 0.669 & 0.769 \\
& (0.017) & (0.016) & (0.016)& (0.014) \\
[3pt]
{$\varphi_{\mathrm{ERADE}}$} & 0.589 & 0.665 & 0.666 &
0.764 \\
& (0.019) & (0.019) & (0.019) & (0.018) \\
\hline
\end{tabular*}
\end{table}
%
\begin{table}[b]
\def\arraystretch{0.9}
\tabcolsep=0pt
\caption{Expectation and standard deviation (within brackets) of the
proportion of allocations to $A$ under the uniform distribution
$\mathcal{U}$ with $(\alpha,\boldsymbol{\tau}^t)=(-4,-1,3,3)$, so that
the optimal compound target is $\pi_\omega^*(0,0)=0.242$, $\pi
_\omega
^*(1,0)=0.209$, $\pi_\omega^*(0,1)=0.415$ and $\pi_\omega^*(1,1)=0.585$}
\label{Tabella 5}
\begin{tabular*}{\textwidth}{@{\extracolsep{\fill}}lcccc@{}}
\hline
& $\boldsymbol{\pi_n(0,0)}$ & $\boldsymbol{\pi_n(1,0)}$ & $\boldsymbol{\pi_n(0,1)}$ & $\boldsymbol{\pi_n(1,1)}$ \\
\hline
{$\varphi_{Z}$} & 0.250 & 0.217 & 0.416 & 0.582 \\
& (0.042) & (0.041) & (0.049)& (0.050) \\
[3pt]
{$\varphi_{BAZ_1}$} & 0.244 & 0.211 & 0.412 & 0.585 \\
& (0.024) & (0.022) & (0.024) & (0.026) \\
[3pt]
{$\varphi_{BAZ_2}$} & 0.244 & 0.212 & 0.415 & 0.585 \\
& (0.013) & (0.013) & (0.017)& (0.016) \\
[3pt]
{$\varphi_{\mathrm{ERADE}}$} & 0.251 & 0.217 & 0.417 &
0.584 \\
& (0.017) & (0.016) & (0.018) & (0.019) \\
\hline
\end{tabular*}
\end{table}

\begin{table}
\tabcolsep=0pt
\caption{Expectation and standard deviation (within brackets) of the
proportion of allocations to $A$ under the nonuniform distribution
$\mathcal{NU}$ with $(\alpha,\boldsymbol{\tau}^t)=(1,1,1,1)$, so that
the optimal compound target is $\pi_\omega^*(0,0)=0.578$, $\pi
_\omega
^*(1,0)=0.700$, $\pi_\omega^*(0,1)=0.743$ and $\pi_\omega^*(1,1)=0.646$}
\label{Tabella 6}
\begin{tabular*}{\textwidth}{@{\extracolsep{\fill}}lcccc@{}}
\hline
& $\boldsymbol{\pi_n(0,0)}$ & $\boldsymbol{\pi_n(1,0)}$ & $\boldsymbol{\pi_n(0,1)}$ & $\boldsymbol{\pi_n(1,1)}$ \\
\hline
{$\varphi_{Z}$} & 0.576 & 0.696 & 0.732 & 0.651 \\
& (0.054) & (0.041) & (0.034)& (0.071) \\
[3pt]
{$\varphi_{BAZ_1}$}& 0.577 & 0.699 & 0.739 & 0.646 \\
& (0.026) & (0.025) & (0.024) & (0.028)\\
[3pt]
{$\varphi_{BAZ_2}$}& 0.577 & 0.698 & 0.740 & 0.646 \\
& (0.017) & (0.015) & (0.014)& (0.017) \\
[3pt]
{$\varphi_{\mathrm{ERADE}}$} & 0.576 & 0.694 & 0.738 &
0.640 \\
& (0.021) & (0.018) & (0.014) & (0.030) \\
\hline
\end{tabular*}
\end{table}

\begin{table}[b]
\tabcolsep=0pt
\caption{Expectation and standard deviation (within brackets) of the
proportion of allocations to $A$ under the nonuniform distribution
$\mathcal{NU}$ with $(\alpha,\boldsymbol{\tau}^t)=(-4,-1,3,3)$, so that
the optimal compound target is $\pi_\omega^*(0,0)=0.279$, $\pi
_\omega
^*(1,0)=0.186$, $\pi_\omega^*(0,1)=0.371$ and $\pi_\omega^*(1,1)=0.534$}
\label{Tabella 7}
\begin{tabular*}{\textwidth}{@{\extracolsep{\fill}}lcccc@{}}
\hline
& $\boldsymbol{\pi_n(0,0)}$ & $\boldsymbol{\pi_n(1,0)}$ & $\boldsymbol{\pi_n(0,1)}$ & $\boldsymbol{\pi_n(1,1)}$ \\
\hline
{$\varphi_{Z}$} & 0.284 & 0.197 & 0.377 & 0.539 \\
& (0.050) & (0.041) & (0.035)& (0.073) \\
[3pt]
{$\varphi_{BAZ_1}$}& 0.279 & 0.188 & 0.373 & 0.535 \\
& (0.026) & (0.021) & (0.026) & (0.024) \\
[3pt]
{$\varphi_{BAZ_2}$}& 0.280 & 0.189 & 0.373 & 0.534 \\
& (0.015) & (0.015) & (0.013)& (0.013) \\
[3pt]
{$\varphi_{\mathrm{ERADE}}$}& 0.286 & 0.195 & 0.375 &
0.533 \\
& (0.019) & (0.018) & (0.014) & (0.023) \\
\hline
\end{tabular*}
\end{table}
In general, $\varphi_{Z}$ in (\ref{gfunc}) is characterized by the
strongest variability with respect to the other procedures, since it is
based only on the current estimate of the target, independently on the
actual allocation proportion. Moreover, as theoretically shown in \cite{Zha07},
 the variability of this design increases as the
representativeness of the strata decreases, and this behavior is also
confirmed by $\varphi_{\mathrm{ERADE}}$; see Tables \ref{Tabella 6} and
\ref{Tabella 7}. On the other hand, $\varphi_{BAZ_1}$ and $\varphi
_{BAZ_2}$ tend to balance the variability of the allocation proportions
between the different population strata. Indeed, the standard
deviations of the design in the different patterns are similar, since
the reinforced doubly adaptive biased coin design forces the closeness
to the desired target the more the strata are
under-represented.\vadjust{\goodbreak}

Note that, in general, discontinuous randomization functions perform
better with respect to the continuous ones. This is quite natural
since, when the allocation proportion is around the target---in
particular, for sufficiently large samples, due to the almost sure
convergence---the continuous allocation procedures randomize the
assignment with probability close to the target, while the
discontinuous ones tend to force the allocation in the same way at each
step, even asymptotically.

Moreover, our simulation study points to the fact that $\varphi
_{BAZ_2}$ guarantees always a stable behavior and tends to have better
performances w.r.t. $\varphi_{\mathrm{ERADE}}$, especially for strongly
under-represented strata. For instance, in the case of a
nonuniform
covariate distribution, when $(t_j,w_l)=(1,1)$ the standard deviation
of the allocation proportions under $\varphi_{BAZ_2}$ is almost half of
the $\varphi_{\mathrm{ERADE}}$'s one (see Tables \ref{Tabella 6} and
\ref{Tabella 7}).\vadjust{\goodbreak}

The same conclusions have been observed through further simulations,
omitted here for brevity, with larger sample size.
However, in this paper we decided to present the case of $n=500$ in
order to emphasize the evolution of our procedure, with respect to the
others, especially for strongly under-represented strata (note that the
expected number of patients within stratum $(1,1)$ under the
nonuniform distribution $\mathcal{NU}$ is $50$).

\section{Discussion}\label{s9}
In the context of clinical trials for treatment comparisons, several
different approaches have been proposed in the literature in order to
provide a valid trade-off between ethical concerns and inferential
precision (such as group sequential designs, interim analysis, etc.)
and over the past 20 years there has been a growing stream of
statistical papers on the topic of response-adaptive randomization;
see, for instance, the seminal books of \cite{Hu07,Ros02}.
Within this framework, in this paper we suggest a design strategy that
combines efficiency and ethical gain for responses following a linear
homoscedastic model.
By using a compound optimization approach we derive optimal target
allocations for the treatments that can be implemented via the adoption
of a new class of CARA randomization procedure. Through the proposed
methodology, the optimal compound allocations move away from balance
(i.e., the optimal inferential target) toward the better treatment
adaptively, on the basis of the treatment effects. Since joint balance
implies marginal one, the proposed design strategy is robust with
respect to
possible misspecification of the model in terms of presence or absence
of interactions among
prognostic factors, or between treatments and covariates.
Moreover, the proposed methodology is quite robust and performs well,
even in the case of approximate homoscedasticity of the outcomes
(perhaps after suitable transformations), as also pointed out by \cite{Atk05a,Ros08}. On the other hand, in the case of heteroscedastic
responses and, more generally, for generalized linear models, balance
does not imply efficiency; our approach could still be applied, but
with different inferential criteria (that could be optimized
numerically) and different weight functions. Further research is needed
on this topic.

\begin{appendix}

\section*{Appendix}\label{app}

\subsection{\texorpdfstring{Proof of Proposition \protect\ref{mainprop}}{Proof of Proposition 3.1}}\label{a1}

\mbox{}

C1. For the sake of simplicity in this \hyperref[app]{Appendix}, we will often omit the subscripts indicating the dimensions
of vectors and matrices.
Let $ \widetilde{\mathbf{M}}=n\mathbf{M}$, then
\[
\det \widetilde{\mathbf{M}} = \det \left(
\begin{array}{c@{\quad}c@{\hspace*{6pt}}|@{\hspace*{6pt}}c@{\quad}c}
\displaystyle\sum_{i=1}^n\delta_i & \boldsymbol{\delta
}^{t}\mathbf{F} & 0 & \boldsymbol{0}_{1\times p} \\
\mathbf{F}^{t}\boldsymbol{\delta} & \mathbf{F}^{t}\boldsymbol
{\Delta}
\mathbf{F} & \mathbf{0}_{p\times1} & \mathbf{0}_{p\times p} \\[-6.6pt]\\[-6pt]
\hline\vspace*{-6pt}\\[-7pt]
0 & \mathbf{0}_{1\times p} &\displaystyle n-\sum_{i=1}^n\delta_i & (\mathbf
{1}-\boldsymbol{\delta})^{t}\mathbf{F} \\
\mathbf{0}_{p\times1} & \mathbf{0}_{p\times p} & \mathbf
{F}^{t}(\mathbf{1}-\boldsymbol{\delta}) &\mathbf{F}^{t}(\mathbf
{I}-\boldsymbol{\Delta})\mathbf{F}\\
\end{array}
\right).
\]
Let $\boldsymbol{\Omega}_A=\mathbf{F}^t \boldsymbol{\Delta} \mathbf
{F}$ and
$\boldsymbol{\Omega}_B=\mathbf{F}^t (\mathbf{I}-\boldsymbol{\Delta} )
\mathbf{F}$, and we obtain that
\begin{eqnarray*}
\det \widetilde{\mathbf{M}}&= & \det \boldsymbol{\Omega}_A
\Biggl\{\sum
_{i=1}^n\delta_i-\boldsymbol{\delta
}^{t}\mathbf{F}\boldsymbol{\Omega}_A^{-1} \mathbf{F}^{t}\boldsymbol
{\delta}\Biggr\}\\
&& {}\times\det  \boldsymbol{\Omega}_B \Biggl\{\Biggl(n-\sum_{i=1}^n\delta
_i\Biggr)-(\mathbf{1}-\boldsymbol{\delta})^{t}\mathbf{F} \boldsymbol{\Omega
}_B ^{-1} \mathbf{F}^{t}(\mathbf{1}-\boldsymbol{\delta}) \Biggr\}.
\end{eqnarray*}
Let now $\boldsymbol{\delta}^t \mathbf{F}=(\mathbf{\widetilde
{N}}_{\mathbb{T}}^t, \mathbf{\widetilde{N}}_{\mathbb{W}}^t, \mathbf
{\widetilde{N}}_{\mathbb{T}\otimes\mathbb{W}}^t)$,
where $\mathbf{\widetilde{N}}_{\mathbb{T}}^t=(\widetilde{N}(t_1),
\ldots
, \widetilde{N}(t_J))$, $\mathbf{\widetilde{N}}_{\mathbb
{W}}^t=(\widetilde{N}(w_1),\ldots, \widetilde{N}(w_L))$ with
$\widetilde
{N}(t_j)=\sum_{
l=0}^{L}\widetilde{N}(j,l)$ and $\widetilde{N}(w_l)=\sum
_{j=0}^{J}\widetilde{N}(j,l)$, and
$\mathbf{\widetilde{N}}_{\mathbb{T}\otimes\mathbb
{W}}^t=(\widetilde
{N}(1,1), \ldots, \widetilde{N}(1,L), \ldots, \widetilde{N}(J,1),
\ldots
, \widetilde{N}(J,L))$.
Clearly, $\widetilde{N}(t_j)$ and $\widetilde{N}(w_l)$ are the number
of subjects assigned to treatment $A$ within category $t_j$ of $T$
$(j=0,\ldots, J)$ and $w_l$ of $W$ $(l=0,\ldots, L)$, respectively.
Also,\vspace*{1pt} let $\mathbf{\widetilde{N}}_{t_j \otimes\mathbb
{W}}^t=(\widetilde
{N}(j,1), \ldots, \widetilde{N}(j,L))$ and $\mathbf{\widetilde
{N}}_{\mathbb{T} \otimes w_l}^t=(\widetilde{N}(1,l), \ldots,
\widetilde
{N}(J,l))$.
Then, the matrix $\boldsymbol{\Omega}_A$ can be partitioned as follows:
%
\begin{equation}\label{pertitomega}
\boldsymbol{\Omega}_A=\left(
\begin{array} {c@{\hspace*{6pt}}|@{\hspace*{6pt}}c}
\mathbf{A} & \mathbf{B}\\
[-6.6pt]\\[-6pt] \hline\vspace*{-6pt}\\[-7pt]
\mathbf{B}^t & \mathbf{C}
\end{array}
\right),
\end{equation}
where
\begin{eqnarray*}
\mathbf{A}&=&\diag(\mathbf{\widetilde{N}}_{\mathbb{T}}, \mathbf
{\widetilde{N}}_{\mathbb{W}})+ \left(
\begin{array} {c@{\quad}c@{\quad}c@{\hspace*{6pt}}|@{\hspace*{6pt}}c}
&&& \mathbf{\widetilde{N}}_{t_1 \otimes\mathbb{W}}^t\\
&\mathbf{0}_{J\times J}& & \vdots\\
&& & \mathbf{\widetilde{N}}_{t_J \otimes\mathbb{W}} ^t\\
[-6.6pt]\\[-6pt] \hline\vspace*{-6pt}\\[-6pt]
\mathbf{\widetilde{N}}_{t_1 \otimes\mathbb{W}}& \cdots& \mathbf
{\widetilde{N}}_{t_J \otimes\mathbb{W}} & \mathbf{0}_{L\times L}
\end{array}
\right),\\
\mathbf{B}&=&\left(
\begin{array} {c@{\quad}c@{\quad}c@{\quad}c@{\quad}c}
\mathbf{\widetilde{N}}_{t_1 \otimes\mathbb{W}}^t & \mathbf{0}&
\mathbf{0} & \cdots& \mathbf{0}\\
\mathbf{0} &\mathbf{\widetilde{N}}_{t_2 \otimes\mathbb{W}}^t &
\mathbf{0} & \cdots& \mathbf{0}\\
\vdots& & \ddots& & \mathbf{0}\\
\mathbf{0} & \mathbf{0} & \cdots& & \mathbf{\widetilde{N}}_{t_J
\otimes\mathbb{W}}^t\\[3pt]
\diag(\mathbf{\widetilde{N}}_{t_1 \otimes\mathbb{W}}) & \diag
(\mathbf
{\widetilde{N}}_{t_2 \otimes\mathbb{W}}) & \cdots& & \diag(\mathbf
{\widetilde{N}}_{t_J \otimes\mathbb{W}})
\end{array}
\right)
\end{eqnarray*}
and $\mathbf{C}=\diag(\mathbf{\widetilde{N}}_{\mathbb{T} \otimes
\mathbb{W}})$, where $\mathbf{0}_{K \times K}$ is the $K$-dim zero
matrix. Thus,
%
\begin{equation}\label{Omega}
\boldsymbol{\Omega}_A^{-1}=\left(
\begin{array} {c@{\hspace*{6pt}}|@{\hspace*{6pt}}c}
\mathbf{0} & \mathbf{0}\\
[-6.6pt]\\[-7pt] \hline\vspace*{-7pt}\\[-7.6pt]
\mathbf{0} & \mathbf{C}^{-1}
\end{array}
\right)+
\left(
\begin{array} {c}
\mathbf{I}_{J+L}\\
-\mathbf{C}^{-1}\mathbf{B}^t
\end{array}
\right)
\boldsymbol{\Gamma}^{-1}
(\mathbf{I}_{J+L} , -\mathbf{B}
\mathbf{C}^{-1}),
\end{equation}
where $\boldsymbol{\Gamma}=\mathbf{A}-\mathbf{B}\mathbf{C}^{-1}\mathbf{B}^t
=\diag(\mathbf{\widetilde{N}}_{\mathbb{T}\otimes w_0},\mathbf
{\widetilde
{N}}_{t_0 \otimes\mathbb{W}})$. Note that $\boldsymbol{\Omega}_A$ is
nonsingular if and only if
$\mathbf{C}$ and $\boldsymbol{\Gamma}$ are nonsingular and
$\det \boldsymbol{\Omega}_A=\det \mathbf{C} \cdot\det
\boldsymbol{\Gamma}
= \prod_{j,l=1}^{J,L} \widetilde{N}(j,l) \prod_{j=1}^{J} \widetilde
{N}(j,0) \prod_{l=1}^{L}
\widetilde{N}(0,l)$.
From (\ref{Omega}) it follows that
\[
\boldsymbol{\delta
}^{t}\mathbf{F} \boldsymbol{\Omega}_A^{-1}\mathbf{F}^{t}\boldsymbol
{\delta}= \boldsymbol{\delta
}^{t}\mathbf{F} \left(
\begin{array}{c@{\hspace*{6pt}}|@{\hspace*{6pt}}c}
\mathbf{0} & \mathbf{0}\\
[-6.6pt]\\[-6pt] \hline\vspace*{-6pt}\\[-7pt]
\mathbf{0} & \mathbf{C}^{-1}
\end{array}
\right) \mathbf{F}^{t}\boldsymbol{\delta} + \boldsymbol{\delta
}^{t}\mathbf{F}\left(
\begin{array} {c}
\mathbf{I}_{J+L}\\
-\mathbf{C}^{-1}\mathbf{B}^t
\end{array}
\right) \boldsymbol{\Gamma}^{-1}
(\mathbf{I}_{J+L} , -\mathbf{B}
\mathbf{C}^{-1})\mathbf{F}^{t}\boldsymbol{\delta},
\]
where
\[
\boldsymbol{\delta
}^{t}\mathbf{F}\left(
\begin{array} {c@{\hspace*{6pt}}|@{\hspace*{6pt}}c}
\mathbf{0} & \mathbf{0}\\
[-6.6pt]\\[-6pt] \hline\vspace*{-6pt}\\[-7pt]
\mathbf{0} & \mathbf{C}^{-1}
\end{array}
\right)\mathbf{F}^{t}\boldsymbol{\delta} =\sum_{j=1}^{J}\sum_{l=1}^{L}
\widetilde{N}(j,l)
\]
and
\begin{eqnarray*}
\boldsymbol{\delta
}^{t}\mathbf{F}  \left(
\begin{array} {c}
\mathbf{I}_{J+L}\\
-\mathbf{C}^{-1}\mathbf{B}^t
\end{array}
\right) \boldsymbol{\Gamma}^{-1}
(\mathbf{I}_{J+L} , -\mathbf{B}
\mathbf{C}^{-1})\mathbf{F}^{t}\boldsymbol{\delta} = \sum_{j=1}^{J}
\widetilde{N}(j,0) +\sum_{l=1}^{L}
\widetilde{N}(0,l),\vadjust{\goodbreak}
\end{eqnarray*}
since
%
\begin{equation}\label{BCinv}
\mathbf{B}\mathbf{C}^{-1}=\left(
\begin{array} {c @{\quad}c@{\quad} c@{\quad} c}
\mathbf{1}_L^t & \mathbf{0} & \cdots& \mathbf{0}\\
\mathbf{0} & \mathbf{1}_L^t & \cdots& \mathbf{0}\\
\vdots& & \ddots& \vdots\\
\mathbf{0} & \mathbf{0} & \cdots& \mathbf{1}_L^t \\
\mathbf{I}_L & \mathbf{I}_L & \cdots& \mathbf{I}_L \\
\end{array}
\right)
\end{equation}
and
%
\begin{eqnarray}\label{GinvBCinv}\qquad
&&\boldsymbol{\Gamma}^{-1}\mathbf{B}\mathbf{C}^{-1}\nonumber
\\[-8pt]
\\[-8pt]
&& \qquad=\left(
\begin{array} {c@{\quad} c@{\quad} c@{\quad} c}
\widetilde{N}(1,0)^{-1}\mathbf{1}_L^t & \mathbf{0} & \cdots&
\mathbf
{0}\\
\mathbf{0} & \widetilde{N}(2,0)^{-1}\mathbf{1}_L^t &\cdots& \mathbf
{0}\\
\vdots& & \ddots& \vdots\\
\mathbf{0} & & \cdots& \widetilde{N}(J,0)^{-1}\mathbf{1}_L^t \\
\diag(\widetilde{N}_{t_0\otimes\mathbb{W}})^{-1} & \diag
(\widetilde
{N}_{t_0\otimes\mathbb{W}})^{-1} & \cdots& \diag(\widetilde
{N}_{t_0\otimes\mathbb{W}})^{-1} \\
\end{array}
\right).\nonumber
\end{eqnarray}
Therefore,
$\boldsymbol{\delta
}^{t}\mathbf{F} \boldsymbol{\Omega}_A^{-1} \mathbf{F}^{t}\boldsymbol
{\delta
}=\sum_{j=1}^{J}\sum_{l=1}^{L}
\widetilde{N}(j,l)+\sum_{j=1}^{J}
\widetilde{N}(j,0) +\sum_{l=1}^{L}
\widetilde{N}(0,l)$,
and consequently $\sum_{i=1}^n\delta_i-\boldsymbol{\delta
}^{t}\mathbf
{F}(\boldsymbol{\Omega}_A)^{-1} \mathbf{F}^{t}\boldsymbol{\delta
}=\widetilde{N}(0,0)$.
Hence,
\[
\det \widetilde{\mathbf{M}}=\prod_{j=0}^{J}\prod_{l=0}^{L}
N(j,l)^2 \pi(j,l) [ 1-\pi(j,l)]
\]
and thus (\ref{Dopt}) follows directly.

C2. Note that $\widetilde{\mathbf{M}}^{-1}$ can be partitioned as follows:
\[
\widetilde{\mathbf{M}}^{-1}= \left(
\begin{array} {c@{\hspace*{6pt}}|@{\hspace*{6pt}}c}
\widetilde{\mathbf{M}}^{-1}_{11} & \widetilde{\mathbf{M}}^{-1}_{12}\\
[-6.6pt]\\[-6pt] \hline\vspace*{-6pt}\\[-6pt]
(\widetilde{\mathbf{M}}^{-1}_{12})^{t} & \widetilde{\mathbf{M}}^{-1}_{22}
\end{array}
\right),
\]
where $\widetilde{\mathbf{M}}^{-1}_{11}$ is given by
\begin{eqnarray*}
&&\left[\left(
\begin{array}{c@{\quad}c}
\displaystyle\sum_{i=1}^n\delta_i & 0 \\
0 & \displaystyle n-\sum_{i=1}^n\delta_i
\end{array}
\right) -\left(
\begin{array}{c@{\quad}c}
\boldsymbol{\delta
}^{t}\mathbf{F}\boldsymbol{\Omega}_A^{-1}\mathbf{F}^{t}\boldsymbol
{\delta}
& 0 \\
0 & (\mathbf{1}-\boldsymbol{\delta})^{t}\mathbf{F}\boldsymbol{\Omega
}_B^{-1}\mathbf{F}^{t} (\mathbf{1}-\boldsymbol{\delta})
\end{array}
\right) \right]^ {-1}\\
&& \qquad=
\left(
\begin{array}{c@{\quad}c}
N(0,0) \pi(0,0) & 0 \\
0 & N(0,0) [1-\pi(0,0)]
\end{array}
\right)^{-1},
\end{eqnarray*}
\[
\widetilde{\mathbf{M}}^{-1}_{12}= -\widetilde{\mathbf{M}}^{-1}_{11}\left(
\begin{array}{c@{\quad}c}
\boldsymbol{\delta
}^{t}\mathbf{F} & \mathbf{0} \\
\mathbf{0} & (\mathbf{1}-\boldsymbol{\delta})^{t}\mathbf{F}
\end{array}
\right)\left(
\begin{array}{c@{\quad}c}
\boldsymbol{\Omega}_A^{-1} & \mathbf{0} \\
\mathbf{0} & \boldsymbol{\Omega}_B^{-1}
\end{array}
\right),
\]
and
\[
\widetilde{\mathbf{M}}^{-1}_{22}=
\left(
\begin{array}{c@{\quad}c}
\boldsymbol{\Omega}_A^{-1} & \mathbf{0} \\
0 & \boldsymbol{\Omega}_B^{-1}
\end{array}
\right)\left[ \mathbf{I}_{2p} - \left(
\begin{array}{c@{\quad}c}
\mathbf{F}^{t}\boldsymbol{\delta} & \mathbf{0} \\
\mathbf{0} & \mathbf{F}^{t} (\mathbf{1}-\boldsymbol{\delta})
\end{array}
\right) \widetilde{\mathbf{M}}^{-1}_{12}
\right].
\]
Thus, $\det( \mathbf{D}^{t}\mathbf{M}^{-1}\mathbf{D})= \det\mathbf
{M}^{-1}_{22}$, which is given by
\begin{eqnarray*}
\det\boldsymbol{\Omega}_A^{-1} \det\biggl[ \mathbf{I}_{p}+ \frac{\mathbf
{F}^{t}\boldsymbol{\delta}\boldsymbol{\delta}^{t}\mathbf{F}
\boldsymbol
{\Omega}_A^{-1}}{N(0,0) \pi(0,0)} \biggr] \det\boldsymbol{\Omega}_B^{-1}
\det\biggl[
\mathbf{I}_{p}+ \frac{\mathbf{F}^{t}( \mathbf{1}-\boldsymbol
{\delta
}) (\mathbf{1}-\boldsymbol{\delta})^{t}\mathbf{F} \boldsymbol{\Omega
}_B^{-1}}{N(0,0) [1-\pi(0,0)]} \biggr].
\end{eqnarray*}
Applying Sylvester's determinant theorem, we obtain that
\begin{eqnarray*}
 \det\boldsymbol{\Omega}_A^{-1}\det\biggl[ \mathbf{I}_{p}+ \frac{\mathbf
{F}^{t}\boldsymbol{\delta}\boldsymbol{\delta}^{t}\mathbf{F}
\boldsymbol
{\Omega}_A^{-1}}{N(0,0) \pi(0,0)} \biggr]&=& \det\boldsymbol{\Omega}_A^{-1} \biggl( 1+
\frac{\boldsymbol{\delta}^{t}\mathbf{F} \boldsymbol{\Omega
}_A^{-1}\mathbf
{F}^{t}\boldsymbol{\delta}}{N(0,0) \pi(0,0)} \biggr)\\
&=& \frac{\sum_{i=1}^n\delta_i} { \prod_{j=0}^{J}\prod_{l=0}^{L} N(j,l)
\pi(j,l) }.
\end{eqnarray*}
Analogously for treatment $B$,
so that (\ref{C3_crit}) follows easily after simple algebra.

C3--C4. Clearly $\tr( n^{-1}\mathbf{D}^{t}\mathbf{M}^{-1} \mathbf{D})=\tr
\widetilde{\mathbf{M}}^{-1}_{22}$, so criterion C3 is given by
%
\begin{eqnarray}\label{traceC3}
\qquad && \sigma^2 \tr\biggl( \boldsymbol{\Omega}_A^{-1} + \frac{1} {N(0, 0) \pi(0, 0)}
\boldsymbol{\Omega}_A^{-1} \mathbf{F}^{t}\boldsymbol{\delta}
\boldsymbol{\delta}^{t}\mathbf{F}\boldsymbol{\Omega}_A^{-1}\biggr) \nonumber
\\[-8pt]
\\[-8pt]
&&\qquad {}+ \sigma^2 \tr\biggl( \boldsymbol{\Omega}_B^{-1}+ \frac{1}{ N(0, 0) [1-\pi(0,
0) ]} \boldsymbol{\Omega}_B^{-1} \mathbf{F}^{t}( \mathbf{1}-\boldsymbol
{\delta}) (\mathbf{1}-\boldsymbol{\delta})^{t}\mathbf{F} \boldsymbol
{\Omega}_B^{-1}\biggr).\nonumber
\end{eqnarray}
Note that $\tr( n^{-1}\sigma^2\mathbf{E}^{t}
\mathbf{M}^{-1}\mathbf{E})$ coincides with (\ref{traceC3}) and thus
C4 is equal to~C3.
Since
\[
\boldsymbol{\Omega}_A^{-1}=\left(
\begin{array} {c@{\hspace*{6pt}}|@{\hspace*{6pt}}c}
\boldsymbol{\Gamma}^{-1} & -\boldsymbol{\Gamma}^{-1} \mathbf{B}\mathbf
{C}^{-1}\\
[-6.6pt]\\[-6pt] \hline\vspace*{-6pt}\\[-6pt]
-\mathbf{C}^{-1} \mathbf{B}^t \boldsymbol{\Gamma}^{-1} & \mathbf
{C}^{-1}+\mathbf{C}^{-1}\mathbf{B}^t \boldsymbol{\Gamma}^{-1}\mathbf
{B}\mathbf{C}^{-1}
\end{array}
\right),
\]
we obtain $\tr(\boldsymbol{\Omega}_A^{-1})=\tr(\boldsymbol{\Gamma
}^{-1})+\tr
(\mathbf{C}^{-1})+\tr(\mathbf{C}^{-1}\mathbf{B}^t \boldsymbol{\Gamma
}^{-1}\mathbf{B}\mathbf{C}^{-1})$.
From (\ref{BCinv}) and (\ref{GinvBCinv}), it follows that
\[
\tr(\mathbf{C}^{-1}\mathbf{B}^t \boldsymbol{\Gamma}^{-1}\mathbf
{B}\mathbf
{C}^{-1})\!=\!\sum_{j=1}^{J} \sum_{l=1}^{L} \biggl( \frac{1} {\widetilde{N} (j,
0)}\!+\! \frac{1} {\widetilde{N} (0, l)} \biggr)\!=\!\sum_{j=1}^{J} \frac{L}
{\widetilde{N} (j, 0)}\!+\!\sum_{l=1}^{L} \frac{J } {\widetilde{N} (0, l)},
\]
and thus
$\tr(\boldsymbol{\Omega}_A^{-1})$ is
\begin{eqnarray*}
\sum_{j=1}^{J} \sum_{l=1}^{L} \biggl( \frac{1} {\widetilde{N} (0, l)}\!+\!
\frac
{1} {\widetilde{N} (j, 0)}\biggr)\!+\!
\sum_{j=1}^{J} \sum_{l=1}^{L} \biggl( \frac{1} {\widetilde{N} (j, l)}
\biggr)\!+\!\sum
_{j=1}^{J} \sum_{l=1}^{L} \biggl( \frac{1} {\widetilde{N} (0, l)}\! +\!\frac{1}
{\widetilde{N} (j, 0)}\biggr).
\end{eqnarray*}
Moreover, from (\ref{Omega}), (\ref{BCinv}) and (\ref{GinvBCinv}) we obtain
$\boldsymbol{\delta}^{t}\mathbf{F}\boldsymbol{\Omega}_A^{-1}=(\mathbf
{1}^t_{J+L},\mathbf{-1}^t_{J \cdot L} )$ and thus
$\boldsymbol{\Omega}_A^{-1} \mathbf{F}^{t}\boldsymbol{\delta
}\boldsymbol
{\delta}^{t}\mathbf{F}\boldsymbol{\Omega}_A^{-1}$ has unitary diagonal elements,
so that\break $\tr(\boldsymbol{\Omega}_A^{-1} \mathbf{F}^{t}\boldsymbol
{\delta
}\boldsymbol{\delta}^{t}\mathbf{F}\boldsymbol{\Omega}_A^{-1})=J+L+J
\cdot L$.
Analogously for treatment $B$.

\subsection{\texorpdfstring{Proof of Theorem \protect\ref{thconvergence}}{Proof of Theorem 7.1}}\label{a2}
As regards the first statement (\ref{Istat}), at~each stratum
$(t_j,w_l)$ we will prove the convergence of both $\pi_n(j,l)$ and
$\widehat{\pi}_n^*(j,l)$ to the target $\pi^*(j,l)$; for ease of
notation we will often omit the subscript
$(j,l)$ assuming that we are fixing the stratum $(t_j,w_l)$. Let
$\mathfrak{F}_n=\sigma(Y_1,\ldots,Y_n,\delta_1,\ldots,\allowbreak\delta
_n,Z_1,\ldots, Z_n)$
denote the $\sigma$-field representing the history of the trial, with
$\mathfrak{F}_0$ the trivial $\sigma$-field, and $\mathfrak
{G}_n=\sigma
(\mathfrak{F}_n, Z_{n+1})$. Moreover, let $\Delta M_i= [\delta_{i}-
E(\delta_{i}| \mathfrak{G}_{i-1})]\mathbh{1}_{\{Z_i=(t_j,w_l)\} }$,
then $\{\Delta M_i; i\geq1\} $ is a sequence of bounded martingale
differences with $|\Delta M_i |\leq1$ for any $i\geq1$; thus the
sequence $\{M_{n}=\sum_{i=1}^{n}\Delta M_i ; \mathfrak{G}_{n}\}$ is a
martingale with \mbox{$\sum_{k=1}^{n} E[(\Delta M_i)^2 |\mathfrak{G}_{k-1}]
\leq N_n$}.

Let $l_n=\max\{s\dvtx 2m+1 \leq s \leq n, \widetilde{N}_s - N_s \widehat
{\pi}^\ast_{s}\leq0 \}$, with $\max\varnothing=2m$, and note that
\begin{eqnarray*}
\widetilde{N}_n &= &\widetilde{N}_{l_n+1}+ \sum_{k=l_n+2}^{n} \Delta
M_k+ \sum_{k=l_n+2}^{n} E(\delta_{k}| \mathfrak{G}_{k-1}) \mathbh{1}_{\{
Z_k=(t_j,w_l)\} }\\
&\leq& \widetilde{N}_{l_n}+1 +M_n -M_{l_n+1} + \sum_{k=l_n+2}^{n}
\varphi(\pi_{{k-1} };\widehat{\pi}_{k-1}^{\ast}; \widehat{p}_{{k-1}}
) \mathbh{1}_{\{Z_k=(t_j,w_l)\} }\\
&< &\widetilde{N}_{l_n}+1+M_n
-M_{l_n+1} + \sum_{k=l_n+2}^{n} \widehat{\pi}^{\ast}_{k-1} \mathbh{1}_{\{Z_k=(t_j,w_l)\} }\\
&=&\widetilde{N}_{l_n}+1+M_n -M_{l_n+1} + \sum_{k=1}^{n} \widehat{\pi
}^{\ast}_{k-1} \mathbh{1}_{\{Z_k=(t_j,w_l)\} } - \sum_{k=1}^{l_n+1}
\widehat{\pi}^{\ast}_{k-1} \mathbh{1}_{\{Z_k=(t_j,w_l)\} },
\end{eqnarray*}
since for any $i\geq l_n+1$, $\varphi(\pi_{{i} };\widehat{\pi
}_{i}^{\ast}; \widehat{p}_{{i}} ) < \widehat{\pi}_{i}^{\ast}< \pi
_{i} $;
whereas $\widetilde{N}_{l_n}\leq N_{l_n} \widehat{\pi}^\ast_{l_{n}}$
and thus
\begin{eqnarray*}
\widetilde{N}_n - N_{n} \widehat{\pi}^\ast_{n}&\leq& \Biggl( N_{l_n}
\widehat{\pi}^\ast_{l_n} - \sum_{k=1}^{l_n+1} \widehat{\pi}^{\ast
}_{k-1}\mathbh{1}_{\{Z_k=(t_j,w_l)\} } \Biggr)+ M_n -M_{l_n+1} +1\\
&&- \Biggl(N_{n} \widehat{\pi
}^\ast_{n}- \sum_{k=1}^{n} \widehat{\pi}^{\ast}_{k-1}\mathbh{1}_{\{
Z_k=(t_j,w_l)\} } \Biggr).
\end{eqnarray*}
Since $p(j,l)>0$ for each stratum $(t_j,w_l)$, then as $n \rightarrow
\infty$, $N_n \rightarrow\infty$ a.s. and therefore
$N_n^{-1}M_n\rightarrow0$ a.s.; see, for instance, \cite{St74}.
Furthermore, as $n \rightarrow\infty$, $\widehat{\mathbf
{p}}_{n}\rightarrow\mathbf{p} $ a.s. and at least one of the
the number of assignments to the treatments, namely $\widetilde{N}_n$
and $(N_n-\widetilde{N}_n)$, tends to infinity a.s. As showed in~\cite{Hu09}, in any case $\widehat{\boldsymbol{\gamma}}_{n}$
has finite limit so that, from the properties of $\pi^*$, there exists
a $v\in(0,1)$ such that
%
\begin{equation}\label{limit1}
\widehat{\pi}^\ast_{n} \rightarrow v\qquad \mbox{a.s.}
\end{equation}
and so
\[
\widehat{\pi}^\ast_{n}- \frac{ \sum_{k=1}^{n} \widehat{\pi
}^{\ast
}_{k-1} \mathbh{1}_{\{Z_k=(t_j,w_l)\}} }{ \sum_{k=1}^{n} \mathbh{1}_{\{
Z_k=(t_j,w_l)\}}}\rightarrow0 \qquad \mbox{a.s.}
\]
As $n \rightarrow\infty$, then $l_n\rightarrow\infty$ or $\sup_n l_n<
\infty$; in either case,
\[
\widehat{\pi}^\ast_{l_n}\frac{N_{l_n}}{N_n} - \frac{ \sum
_{k=1}^{l_n+1} \widehat{\pi}^{\ast}_{k-1} \mathbh{1}_{\{
Z_k=(t_j,w_l)\}
} }{ \sum_{k=1}^{n} \mathbh{1}_{\{Z_k=(t_j,w_l)\}}}\rightarrow0
\qquad\mbox{a.s.}
\]
and therefore
%
\begin{equation}\label{dim1}
( \pi_n-\widehat{\pi}^\ast_{n} ) ^+ \rightarrow0 \qquad\mbox{a.s.}
\end{equation}
Let $\lambda_n=\max\{s\dvtx 2m+1 \leq s \leq n, (N_s-\widetilde{N}_s) -
N_s(1-\widehat{\pi}^\ast_{s})\leq0 \}$, for any $i\geq\lambda_n+1$,
we have $\varphi(\pi_{i} ;\widehat{\pi}_{i}^{\ast}; \widehat{p}_{{i}}
)=1-\varphi(1-\pi_{i} ;1-\widehat{\pi}_{i}^{\ast}; \widehat{p}_{{i}}
)> \widehat{\pi}_{i}^{\ast}>\pi_{i}$.
Then,
\begin{eqnarray*}
N_n-\widetilde{N}_n &=& N_{\lambda_n+1}-\widetilde{N}_{\lambda_n+1}
+ \sum_{k=\lambda_n+2}^{n} E\bigl((1-\delta_{k})| \mathfrak{G}_{k-1}\bigr)
\mathbh{1}_{\{Z_k=(t_j,w_l)\} } \\
&& {} +  \sum_{k=\lambda_n+2}^{n} \bigl[ (1-\delta_{k})- E\bigl((1-\delta_{k})|
\mathfrak{G}_{k-1}\bigr) \bigr] \mathbh{1}_{\{Z_k=(t_j,w_l)\} } \\
&\leq& N_{\lambda_n}+1-\widetilde{N}_{\lambda_n} -(M_n -M_{\lambda
_n+1}) \\
&&{}+ \sum_{k=\lambda_n+2}^{n} \varphi(1-\pi_{k-1} ;1-\widehat{\pi
}_{k-1}^{\ast}; \widehat{p}_{k-1} ) \mathbh{1}_{\{Z_k=(t_j,w_l)\} }\\
&< & N_{\lambda_n}+1-\widetilde{N}_{\lambda_n}- ({M}_n -{M}_{\lambda
_n+1}) + \sum_{k=\lambda_n+2}^{n} (1-\widehat{\pi}^{\ast}_{k-1})
\mathbh{1}_{\{Z_k=(t_j,w_l)\} }\\
&= & N_{\lambda_n}+1-\widetilde{N}_{\lambda_n}-({M}_n -{M}_{\lambda
_n+1}) + \sum_{k=1}^{n} (1-\widehat{\pi}^{\ast}_{k-1})\mathbh{1}_{\{
Z_k=(t_j,w_l)\} }\\
&& {} - \sum_{k=1}^{\lambda_n+1} (1-\widehat{\pi
}^{\ast
}_{k-1})\mathbh{1}_{\{Z_k=(t_j,w_l)\} }.
\end{eqnarray*}
Hence,
\begin{eqnarray*}
&&(N_n-\widetilde{N}_n)-N_n(1- \widehat{\pi}^\ast_{n})\\
&&\qquad \leq\Biggl\{N_{\lambda_n}(1-\widehat{\pi}_{\lambda_n}^{\ast}) - \sum_{k=1}^{\lambda_n+1}
(1-\widehat{\pi}^{\ast}_{k-1}) \mathbh{1}_{\{Z_k=(t_j,w_l)\} }\Biggr\}\\
&&\hphantom{\qquad \leq} {}+1-({M}_n - {M}_{\lambda_n+1}) -\Biggl\{N_n(1- \widehat{\pi}^\ast
_{n})-\sum
_{k=1}^{n} (1-\widehat{\pi}^{\ast}_{k-1}) \mathbh{1}_{\{
Z_k=(t_j,w_l)\} } \Biggr\},
\end{eqnarray*}
so that
%
\begin{equation}\label{dim2}
\bigl( (1-\pi_n)- (1-\widehat{\pi}^\ast_{n}) \bigr) ^+ \rightarrow0 \qquad\mbox{a.s.}
\end{equation}
From (\ref{dim1}) and (\ref{dim2}), as $n$ tends to infinity
\[
\pi_n-\widehat{\pi}^\ast_{n} \rightarrow0\qquad \mbox{a.s.}
\]
and by (\ref{limit1})
\[
\lim_{n\rightarrow\infty} {\pi_{n}}=\lim_{n\rightarrow\infty}
\widehat
{\pi^\ast_{n}} =v\qquad \mbox{a.s.}
\]
Since $0< v <1$, then $0< 1-v <1$ and thus
\[
\lim_{n\rightarrow\infty} \widetilde{N}_n \rightarrow\infty\qquad\mbox
{a.s.} \quad\mbox{and} \quad\lim_{n\rightarrow\infty}(N_n-\widetilde{N}_n)
\rightarrow\infty\qquad\mbox{a.s.}
\]
Therefore, $\lim_{n\rightarrow\infty}\widehat{\boldsymbol{\gamma
}}_{n}\!\rightarrow\!\boldsymbol{\gamma}$ a.s. and from the continuity of
the target\break $\lim_{n\rightarrow\infty} \widehat{\pi}^\ast_{n}=\pi^*=v$
a.s., that is,
%
\begin{equation}\label{strng cons}
\lim_{n\rightarrow\infty} \pi_n=\pi^* \qquad \mbox{a.s.}
\end{equation}
Taking into account the average information matrix $\mathbf{M}$ in
(\ref{MconInter}), from (\ref{strng cons}) and the proof of Proposition
\ref{mainprop} it follows that
\begin{eqnarray*}
\displaystyle\lim_{n\rightarrow\infty} n^{-1}\sum_{i=1}^{n} \delta_i&=& \sum
_{j=0}^{J}\sum_{l=0}^{L}\pi^*(j,l)p(j,l)\qquad  \mbox{a.s.}\\
\displaystyle\lim_{n\rightarrow\infty}
n^{-1}\boldsymbol{\delta}^{t}\mathbf{F}&=&
\widetilde{{\boldsymbol{\pi}}}^{* t}\qquad \mbox{a.s.}\quad \mbox{and}\quad \lim
_{n\rightarrow\infty} n^{-1}\mathbf{F}^{t}\boldsymbol{\Delta
}\mathbf
{F} = \boldsymbol{\mathcal{M}}^A\quad \mbox{a.s.}
\end{eqnarray*}
Thus, as $n$ goes to infinity the information matrix converges almost
surely to $\boldsymbol{\mathcal{M}}(\boldsymbol{\pi}^{\ast})$,
which is
nonsingular since $\sum_{j=0}^{J}\sum_{l=0}^{L}\pi^*(j,l)p(j,l)\in
(0,1)$, $\boldsymbol{\mathcal{M}}^A$ and $\boldsymbol{\mathcal{M}}^B$
are nonsingular, and the matrix
\begin{eqnarray*}
&&\left(
\begin{array} {c@{\quad} c}
\displaystyle\sum_{j=0}^{J}\sum_{l=0}^{L}\pi^*(j,l)p(j,l) & 0 \\
0 &\displaystyle 1-\sum_{j=0}^{J}\sum_{l=0}^{L} \pi^*(j,l)p(j,l) \\
\end{array}
\right)\\
&&\quad -\left(
\begin{array} {c @{\quad}c}
\boldsymbol{\delta}^{t}\mathbf{F} & \mathbf{0} \\
\mathbf{0} & (\mathbf{1}-\boldsymbol{\delta})^{t}\mathbf{F} \\
\end{array}
\right)
\left(
\begin{array} {c @{\quad}c}
\boldsymbol{\mathcal{M}}^A & \mathbf{0} \\
\mathbf{0} & \boldsymbol{\mathcal{M}}^B \\
\end{array}
\right)^{-1}
\left(
\begin{array} {c@{\quad} c}
\mathbf{F}^{t}\boldsymbol{\delta} & \mathbf{0} \\
\mathbf{0} & \mathbf{F}^{t}(\mathbf{1}-\boldsymbol{\delta}) \\
\end{array}
\right)\\
&&\qquad=\left(
\begin{array} {c@{\quad} c}
\pi^*(0,0)p(0,0) & 0 \\
0 & 1-\pi^*(0,0)p(0,0) \\
\end{array}
\right)
\end{eqnarray*}
is nonsingular too, since $\pi^*(0,0)p(0,0)\in(0,1)$. Thus, the
asymptotic normality of $\widehat{\boldsymbol{\gamma}}_{n}$ follows
directly.
\end{appendix}

\section*{Acknowledgments}
The authors thank the referees and the Associate Editor for their
helpful comments, which led to a substantial improved version of the paper.

\begin{supplement}
\stitle{Supplement to ``Multi-objective optimal designs in comparative
clinical trials with covariates: the reinforced doubly adaptive biased
coin design''}
\slink[doi]{10.1214/12-AOS1007SUPP} 
\sdatatype{.pdf}
\sfilename{aos1007\_supp.pdf}
\sdescription{An online supplementary file contains the extension of
inferential criteria C1--C5 to the case of several
covariates.}
\end{supplement}


%
\printaddresses

\end{document}